\documentclass[12pt]{article}
\pdfoutput=1
\usepackage{jheppub,amsmath,amssymb,amsfonts,amsxtra, mathrsfs, makeidx,graphics,graphicx,amsthm,epsfig, youngtab,bm,longtable,float}
\newcommand{\ba}{\begin{array}}
\newcommand{\ea}{\end{array}}
\newcommand{\bi}{\begin{itemize}}
\newcommand{\ei}{\end{itemize}}
\def\vec#1{\bm{#1}}
\def\bea#1\eea{\allowdisplaybreaks \begin{align}#1\end{align}}
 \newcommand{\ben}{\begin{enumerate}}
\newcommand{\een}{\end{enumerate}}
\newcommand{\bean}{\begin{eqnarray*}}
\newcommand{\eean}{\end{eqnarray*}}
\newcommand{\eref}[1]{(\ref{#1})}

\newcommand{\tref}[1]{Table~\ref{#1}}

\newcommand{\nn}{\nonumber}

\newcommand{\PE}{\mathop{\rm PE}}

\newcommand{\BC}{\mathbb{C}}

\newcommand{\BZ}{\mathbb{Z}}

\newcommand{\comment}[1]{}

\newcommand{\CM}{{\cal M}}

\newcommand{\CN}{{\cal N}}

\newcommand{\CR}{{\cal R}}
\newcommand{\CI}{{\cal I}}

\newcommand{\vect}{\mathrm{vector}}
\newcommand{\adj}{\mathbf{Adj}}

\newcommand{\ie}{{\it i.e.}}
\newcommand{\eg}{{\it e.g.}}

\newcommand{\ud}{\mathrm{d}}

\title{5d gauge theories on orbifolds and 4d 't Hooft line indices}

\author[a,b]{Noppadol Mekareeya}
\author[c]{and Diego Rodr\'iguez-G\'omez}

\affiliation[a]{Max-Planck-Institut f\"ur Physik 
(Werner-Heisenberg-Institut),\\
F\"ohringer Ring 6, 80805 M\"unchen, Deutschland}
\affiliation[b]{
Theory Group, Physics Department, CERN, CH-1211, Geneva 23, Switzerland
}
\affiliation[c]{Department of Physics, 
Universidad de Oviedo \\
Avda. Calvo Sotelo 18, 33007, Oviedo, Spain}

\emailAdd{noppadol@mpp.mpg.de, d.rodriguez.gomez@uniovi.es}

\preprint{
\begin{flushright}
MPP-2013-244\\
\end{flushright}
}

\abstract{We study indices for 5d gauge theories on $S^1\times S^4/\mathbb{Z}_n$. In the large orbifold limit, $n\rightarrow\infty$, we find evidence that the indices become 4d indices in the presence of a 't Hooft line operator. The non-perturbative part of the index poses some subtleties when being compared to the 4d monopole bubbling which happens in the presence of 't Hooft line operators. We study such monopole bubbling indices and find an interesting connection to the Hilbert series of the moduli space of instantons on an auxiliary ALE space.
\\
\\
\today
}

\begin{document}
\maketitle

\section{Introduction}

The case of 5d gauge theories has been poorly studied, at least compared to other dimensionalities. It is therefore interesting to study their relatively unexplored landscape. Moreover, 5d gauge theories lie in between of the most familiar and well understood case of 4d gauge theories and the mysterious 6d $(2,\,0)$ CFT, thus potentailly incorporating features of the latter accessible by means of the well-understood techniques developed for the former. Indeed, over the very recent past we have seen quite a lot of developments along this direction \cite{Kallen:2012cs, Kallen:2012va, Kallen:2012zn, Minahan:2013jwa, Qiu:2013pta, Kim:2012qf, Kim:2012tr, Kim:2013nva}. Furthermore, through dimensional reduction interesting connections among theories in different dimensions emanating from 6d and passing through the 5d case have been very recently developed \cite{Cordova:2013cea, Cordova:2013bea}.

On the other hand, 5d gauge theories are interesting by themselves. In particular, they can be at fixed points with rather remarkable properties such as enhanced exceptional global symmetries \cite{Seiberg:1996bd, Morrison:1996xf,Intriligator:1997pq}; see also \cite{Kim:2012gu, Iqbal:2012xm}. Moreover, for some of those CFT's, the gravity dual for the large $N$ limit has been found \cite{Brandhuber:1999np, Bergman:2012kr} (see also \cite{Passias:2012vp,Lozano:2012au}) and quite non-trivial tests of the duality have been performed \cite{Jafferis:2012iv, Bergman:2012qh,Assel:2012nf, Bergman:2013koa}.

Very recently, a very powerful set of exact techniques have been developed to study gauge theories in diverse dimensions. By using the power of localization, partition functions and indices for a very wide variety of theories in different dimensions have been computed. In this paper we will concentrate on indices for 5d gauge theories. Although indices are, in a sense, very coarse observables as only very particular and protected operators contribute, they provide very solid information. We can however have more refined information by putting the theory on more complicated backgrounds. As the index will be sensitive to the background geometry, computing indices in a variety of spaces leads to a deeper understanding of the theory. In particular, global properties of the gauge group which determine the set of allowed line defects \cite{Aharony:2013hda} are expected to emerge in a manifest way as the theory is placed in a non-trivial background \cite{Razamat:2013opa}. In this paper, following this strategy, we will compute indices for gauge theories on orbifolds. 

More precisely, we will consider gauge theories on $S^1\times S^4/\mathbb{Z}_n$, which is conformally equivalent to the compactification of $\mathbb{R}\times \mathbb{C}^2/\mathbb{Z}_n$. Note that $\pi_1(S^3/\mathbb{Z}_n)=\mathbb{Z}_n$, and so there can be a non-trivial monodromy of the gauge field. This is similar in spirit to the so-called lens space index for 4d gauge theories recently considered in \cite{Benini:2011nc, Razamat:2013opa}. 

The unorbifolded, $n=1$ case has been studied in \cite{Kim:2012gu}, where it was shown that the index can be computed as an integral over the gauge holonomies with the appropriate Haar measure of a function containing a perturbative factor with the plethystic exponential of a single-letter index and a non-perturbative factor which coincides with the Nekrasov instanton partition function. In this paper we will extend these results to the general orbifold case. While the structure of the index will be analogous to the unorbifolded case, we need to determine the effect of the orbifold on each term. This requires to specify the degree of the orbifold $n$ as well as its action on both spacetime and gauge fugacities. The latter is determined by the choice of a vector $\vec{r}$ of weights of the gauge fugacities which encodes the monodromy of the gauge field.

Since the background geometry contains two circles, namely the orbifolded one and the ``time" $S^1$, it is natural to consider reductions of the index for a given theory along them. Reducing along the ``time" $S^1$ produces the partition function of the 4d version of the theory on $S^4/\mathbb{Z}_n$. On the other hand, we will find evidence that reduction along the orbifolded $S^1$, implemented by taking the large orbifold limit $n\rightarrow \infty$, leads to the 4d 't Hooft index of \cite{Ito:2011ea, Gang:2012yr}. 

The structure of this paper is as follows. In section \ref{orbifoldindex} we describe the salient features of 5d gauge theories and the computation of their index when placed on the orbifold geometry. It is then easy to see that the reduction along the ``time" $S^1$ immediately recovers the 4d results for the 4d partition function on an ALE space. In section \ref{largeorbifold} we perform the large orbifold reduction and show evidence that we recover the 't Hooft line index. Inspired by this result, in section \ref{monopolebubbling} we study the 4d 't Hooft line index, focusing on the nonperturbative contribution due to monopole bubbling. Interestingly, we find that for a given monopole and a given bubbling, the bubbling index is computed by the Hilbert series of the moduli space of an instanton specified by the chosen monopole and bubbling, this along the lines of Kronheimer's correspondence between instantons and monopoles \cite{kronheimer1985monopoles}. In section \ref{conclusions} we summarize our results and discuss open issues, in particular the fate of monopole bubbling as in 5d. Finally, we postpone to appendix \ref{largeB} some explicit results for monopole bubblings in pure $U(N)$ gauge theory.

\section{Indices for 5d gauge theories on orbifolds}\label{orbifoldindex}

In 5d the minimal supersymmetry contains 8 supercharges. The basic building blocks for the theories of interest are the vector multiplet --containing the gauge field, a real scalar and a symplectic-Majorana gaugino-- and the hypermultiplet --containing 4 real scalars and a complex Dirac fermion--. One salient feature of gauge theories in 5d is that, in addition to other possible global currents, there is a topologically conserved global current $j=\star\,{\rm Tr}\,F\wedge F$ associated to each vector multiplet. The electrically charged excitations are particle-like solitons with instanton charge in a codimension 1 submanifold. These particles are usually called instanton particles and the topologically conserved current instanton current. This current can be gauged by adding a Chern-Simons term to the action $ \int A\wedge F\wedge F$. Note that the 5d Chern-Simons term, being cubic, is proportional to the third order Casimir of the gauge group, and hence automatically vanishes for $USp$ groups. It is also worth mentioning that the effective action for 5d gauge theories on their Coulomb branch can be exactly computed, as it follows from a prepotential severely constrained by gauge invariance. In addition, a similar effect to the 3d parity anomaly whereby upon integrating out a massive Dirac fermion a $\frac{{\rm sign}(m)}{2}$ shift of the Chern-Simons coefficient is produced, also plays a key role in determining the exact prepotential on the Coulomb branch. We refer to \cite{Seiberg:1996bd, Morrison:1996xf,Intriligator:1997pq} for further details on the dynamics of 5d gauge theories.

In order to compute the index for the 5d theories, one considers the Euclidean theory in radial quantization, which amounts to put it on $S^1\times S^4/\mathbb{Z}_n$. More explicitly, we consider a 5d gauge theory on (euclidean) $\mathbb{R}\times\mathbb{C}^2/\mathbb{Z}_n$. Introducing complex coordinates $(z_1,\,z_2)$ on $\mathbb{C}^2$, the orbifold will act as
\begin{equation}
(z_1,\,z_2)\,\sim\,(\omega\,z_1,\,\omega^{-1}\,z_2)\, ,\qquad \omega^n=1\, .
\end{equation}

Note that the cases $n=1,\,2$ are special, since they preserve $SU(2)\times SU(2)$, while for $n>2$ the symmetry is $U(1)\times SU(2)$. Besides, all supercharges are preserved by this orbifold action.

Writting the $\mathbb{R}\times \mathbb{C}^2/\mathbb{Z}_n$ space metric as $ds^2=dx_0^2+dr^2+r^2\,d\Omega_{S^3/\mathbb{Z}_n}^2$ --here $d\Omega_{S^3/\mathbb{Z}_n}^2$ is the standard metric on the lens space-- and upon defining $x_0=e^{-\tau}\,\cos\alpha$ and $r=e^{-\tau}\,\sin\alpha,\,\alpha\in [0,\,\pi]$ and compactifying $\tau$ into an $S^1$, the metric becomes conformally equivalent that of $S^1\times S^4/\mathbb{Z}_n$. One then chooses a supercharge $Q$ and its complex conjugate, so that only primary operators annihilated by this subalgebra contribute to the index weighted by their representation under all other commuting charges. Starting by the unorbifolded case, in 5d the bosonic part of the $\mathcal{N}=1$ superconformal algebra is $SO(2,\,5)\times SU(2)_R$, where $SU(2)_R$ is the R-symmetry. In turn $SO(2,\,5)$ contains the dilatation operator as well as a compact $SO(5)_L$ acting on the $S^4$. The maximal compact subgroup is $[SU(2)_1\times SU(2)_2]_L\times SU(2)_R$. Calling the $U(1)$ Cartans respectively $j_1,\,j_2,\,R$, the generators commuting with the chosen supercharge are $j_2$ and $j_1+R$. Then, the index reads \cite{Bhattacharya:2008zy, Kim:2012gu}

\begin{equation}
\label{index}
\mathcal{I}={\rm Tr}\,(-1)^F\,e^{-\beta\,\Delta}\,x^{2\,(j_1+R)}\,y^{2\,j_2}\,\mathfrak{q}^{\mathfrak{Q}}\,,\qquad \Delta=\epsilon_0-2\,j_1-3\,R\,,
\end{equation}
where $\mathfrak{Q}$ collectively stands for all other commuting global symmetries --including the instanton current-- with associated fugacities collectively denoted by $\mathfrak{q}$. As the index does not depend on $\beta$, only states whose scaling dimension satisfies $\epsilon_0=2\,j_1+3\,R$ contribute. In \cite{Kim:2012gu} it was shown that the index admits a path integral representation obtained by computing the supersymmetric partition function with the appropriate boundary conditions for fermions upon adding chemical potentials for the global symmetries. This partition function is technically computed by adding a $Q$-exact term to the action, which has the effect of localizing the theory on the saddle points of this $Q$-deformed action. As shown in  \cite{Kim:2012gu}, the final result for the index is

\begin{equation}
\label{indexstructure}
\mathcal{I}=\int [\ud\alpha]\,\mathcal{I}_{{\rm p}}\,\mathcal{I}_{{\rm inst}}\,,
\end{equation}
where $\int [\ud\alpha]$ stands for the integration over the gauge group with the suitable Haar measure, while $\mathcal{I}_{p}$ and $\mathcal{I}_{\rm inst}$ stand respectively for the perturbative and instantonic contributions to the index. The perturbative contribution can be thought as the plethistic exponential of the single-letter indices associated to each multiplet present in the theory, that is, schematically
\begin{equation}
\mathcal{I}_{\rm p}={\rm PE} \left[\sum_{V\in \text{vectors}} f_{\vect}^V+\sum_{H\in \text{hypers}} f_{\rm matter}^H \right]
\end{equation}
being $f_{\vect}$, $f_{\rm matter}$ the single-letter contributions to the index. In the cohomological formulation, such single-letter indices are basically given by the Atiyah-Singer index of the appropriate complex depending on the type of multiplet \cite{Pestun:2007rz,Kim:2012gu} (see sect. \ref{orbifoldindex} for explicit expressions). In turn, the instanton part is associated with instantonic particles and it coincides with the 5d Nekrasov instanton partition function.

In the orbifolded case $n\geq 2$ the $[SU(2)_1\times SU(2)_2]_L$ Lorentz symmetry is generically reduced to $[U(1)\times SU(2)]_L$. Nevertheless the localization computation is otherwise exactly analogous to the $n=1$ case. Hence the structure of the index is exactly the same as in the unorbifolded case, with the only difference that both the perturbative and non-perturbative parts must be computed on the orbifold background. As for the perturbative part, the single-letter contributions are given by the indices of the corresponding complexes on the orbifold, which can simply be computed by projecting the unorbifolded case to orbifold-invariants. In turn, as for the non-perturbative contribution, we should compute the Nekrasov instanton partition function on the orbifold geometry.

In the following we will concentrate on $U(N)$ gauge theories. It is therefore useful to recapitulate the most salient features of the topological classification of $U(N)$ bundles on ALE spaces (see \cite{DHMRS2013} and references therein for a more thorough review). A $U(N)$ bundle on $\mathbb{C}^2/\mathbb{Z}_n$ is topologically classified by $n-1$ first Chern classes and one second Chern class. In addition, since $\pi_1(S^3/\mathbb{Z}_n)=\mathbb{Z}_n$, we need to specify the monodromy of the gauge field, labelled by a partition of $N$ as $\vec{N}=(N_1,\,\cdots,\,N_n)$ such that $\sum N_i=N$.

\subsection{Perturbative contribution}

The perturbative contribution to the index of the vector multiplet and hypermultiplet can be read, respectively, from the self-dual complex and the Dirac complex \cite{Kim:2012gu}. The respective contributions can be easily obtained by first computing the equivariant index of the corresponding complex and then taking its plethystic exponential.

The relevant complexes will be the self-dual complex --related to the vector multiplet contribution-- and the Dirac complex --related to the hypermultiplet contribution--. As we are interested on 5d gauge theories on $\mathbb{C}^2$ they will depend on two spacetime fugacities $t_1,\,t_2$ associated to the two $\mathbb{C}$ planes. The relation of these to the more standard $\{x,\,y\}$ used in \cite{Kim:2012qf} is simply 
\bea
t_1=xy,  \qquad t_2=x y^{-1}~,
\eea  
where $x$ and $y$ are fugacities for $U(1) \times SU(2)$ isometry of $\BC^2$ appearing in (\ref{index}).

The action of the orbifold on the Lorentz fugacities is simply
\begin{equation}
t_1 \rightarrow \omega\,t_1\qquad t_2\rightarrow \omega^{-1}\,t_2~, \label{actiont}
\end{equation}
Besides, let us call  gauge symmetry fugacities by $z_\alpha$.\footnote{Note that some of these fugacities might in the end not be gauged and thus correspond to global symmetries. As an example, suppose a $U(N)$ versus an $SU(N)$ gauge theory, whose difference is the overall $U(1)$ being either a gauge symmetry or a global baryonic symmetry. We can, nevertheless, think of all fugacites as gauged ones and decide wether to actually gauge them or not only at the end when integrating over them or not.}  The orbifold will generically have a non-trivial action also on them. Let us particularize now to the $U(N)$ case, where we have
\begin{equation}
z_\alpha\rightarrow \omega^{r_\alpha}\,z_\alpha~, \label{actionz}
\end{equation}
where $\alpha=1, \ldots, N$ and $0 \leq r_\alpha \leq n-1$ for all $\alpha$. In fact, the $r_\alpha$ are related to the monodromy of the orbifold action on the gauge bundle $\vec{N}=(N_1,\cdots,\,N_n)$ as
\begin{equation}
N_{i}=\sum_{\alpha=1}^{N}\,\delta_{r_\alpha, i \,({\rm mod}\,n)}\, ,\qquad i=1,\cdots,\,n
\end{equation}
where $\alpha~ ({\rm mod}~ n)$ runs over $0, \ldots, n- 1$. Therefore, $N_i$ is the number of times that
$i~ ({\rm mod}~ n)$ appears in the vector $\vec r$. 
If we are interested on $SU(N)$, since $\prod_{\alpha=1}^N \,z_\alpha=1$, we must impose 
\bea
\sum_{\alpha=1}^N\,r_\alpha=0~, \qquad \text{for $SU(N)$}~.  \label{constr}
\eea

For example, for $SU(2)$ gauge theory on $\mathbb{C}^2/\mathbb{Z}_2$, corresponding to $(N,n)=(2,2)$, the possibilities are
\bea
\begin{array}{llll}
\vec r= (0,0) \quad &\Rightarrow \quad \vec N = (0,2) ~,  \\
\vec r= (1,1)  \quad &\Rightarrow \quad \vec N = (2,0)~. \label{C2Z2}
\end{array}
\eea
A more extensive list is given in eqs. (2.65)--(2.70) in \cite{DHMRS2013}.

As shown in \cite{Kim:2012qf}, the contribution of each type of multiplet is related to the Atiyah-Singer index of a certain complex,  denoted generically by ${\rm ind}[D(\mathbb{C}^2)](t_1,\,t_2,\,z_\alpha)$. Thus, the index for the complex upon performing the orbifold projection can be done by implementing such projection on the $\mathbb{C}^2$ index. Explicitly
\begin{equation}
\label{prescription}
{\rm ind}[D(\mathbb{C}^2/\mathbb{Z}_2)](t_1,\,t_2,\,z_\alpha)=\frac{1}{n}\,\sum_{j=0}^{n-1}\,{\rm ind}[D(\mathbb{C}^2](\omega^j\,t_1,\,\omega^{-j}\,t_2,\,\omega^{j\,r_\alpha}\,z_\alpha)\, .
\end{equation}

\subsubsection{Vector multiplet contribution}
The relevant complex for the vector multiplet is the self-dual complex. Let us borrow the result for the unorbifolded case from  \cite{Kim:2012gu} for the equivariant index of the self-dual complex (we strip off gauge fugacities)
\begin{equation}
\label{SDindex}
{\rm ind}[D_{SD}(\mathbb{C}^2)](t_1, t_2)=\frac{1+t_1\,t_2}{(1-t_1)\,(1-t_2)}\, .
\end{equation}
Denoting the gauge holonomies $\alpha_i$ and the adjoint character by $\chi_{\rm\mathbf{\adj}}$, the contribution to the index of the vector multiplet is then $\int \prod d\alpha_i\,{\rm PE}[{\rm ind}[-D_{SD}(\mathbb{C}^2)]\,\chi_{\rm \mathbf{\adj}}]$. Following \cite{Kim:2012gu}, the integrand can be manipulated as follows
\begin{equation}
{\rm PE}[{\rm ind}[-D_{SD}(\mathbb{C}^2)]\,\chi_{\rm \mathbf{\adj}}]={\rm PE}[\chi_{\rm\mathbf{\adj}}-{\rm ind}[D_{SD}]]\,{\rm PE}[-\chi_{\rm \mathbf{\adj}}]
\end{equation}
so that $\int \prod d\alpha_i\,{\rm PE}[-\chi_{\rm \mathbf{\adj}}]=\int [d\alpha]$ becomes the gauge group integration with the Haar measure, effectively leaving the contribution of the vector multiplet 
\bea
H^{\text{1-loop}, \BC^2}_{\text{vector}} (t_1, t_2,\vec z) =  {\PE}[f^{\BC^2}_{\vect}(t_1,t_2, \vec z)]
\eea
with
\bea\label{fC2vec}
f^{\BC^2}_{\vect} (t_1, t_2, \vec z) &=(1-{\rm ind}[D_{SD}])\,\chi_{\rm\mathbf{\adj}}(\vec z) = -\frac{t_1 +t_2}{(1-t_1)\,(1-t_2)}\,\chi_{\rm\mathbf{\adj}} (\vec z)~.
\eea
Writting this in terms of the $x,\,y$ one recovers the vector multiplet contribution in \cite{Kim:2012gu}.

\subsubsection*{Reduction on $S^1$} 
Before proceeding further, let us point out that we can reduce this part $H^{\BC^2}_{\text{vector}} (t_1, t_2,\vec z)$ of partition function on the ``time'' circle $S^1$.  This reproduces the well-known formula for one-loop contribution of the vector multiplet \cite{Nekrasov:2003rj} (see also, \eg~ (B.21) of \cite{Alday:2009aq}).  Let us denote by $\beta$ the radius of the circle $S^1$.  The variables $t_{1,2}$ are related to the $\Omega$-deformation parameters $\epsilon_{1,2}$ and gauge parameters $a_\alpha$ (with $\alpha =1,\ldots, N$) as follows:
\bea \label{Nekvars}
t_1 = x y = e^{-\beta \epsilon_1}~, \qquad t_2 = x y^{-1} = e^{-\beta \epsilon_2}~, \qquad z_\alpha = e^{-\beta a_\alpha}~.
\eea
and let us then focus on the limit $\beta \rightarrow 0$.  

Let us consider $U(N)$ gauge group.  From \eref{fC2vec}, we obtain
\bea
H^{\text{1-loop},\BC^2}_{\text{vector}} (t_1, t_2,\vec z) 
&= \PE \left[- \frac{t_1+t_2}{(1-t_1)(1-t_2)}\chi_{\adj}( \vec z) \right] \nn \\
&= \PE \left[\left( \sum_{1\leq i, j \leq N} z_\alpha z_\beta^{-1} \right)\left( \sum_{m,n \geq 1}t_1^{m}t_2^{-n}  + t_1^{(m-1)}t_2^{-(n-1)} \right)  \right] \nn \\
&=  \prod_{m,n \geq 1}  \frac{1}{\prod_{1\leq i ,j \leq N} \{1-t_1^{m}t_2^{-n} z_\alpha z_\beta^{-1} \}\{1-t_1^{(m-1)}t_2^{-(n-1)}z_\alpha z_\beta^{-1} \} }~, \label{5d1loopvector} 
\eea 
Substituting into it \eref{Nekvars} and taking limit $\beta \rightarrow 0$, we obtain
\bea
1-t_1^{m}t_2^{-n} z_\alpha z_\beta^{-1}  \quad &\rightarrow \quad m\epsilon_{1}- n \epsilon_2+ a_\alpha -a_\beta \nn \\
1-t_1^{(m-1)}t_2^{-(n-1)}z_\alpha z_\beta^{-1}\quad &\rightarrow \quad (m-1)\epsilon_{1}- (n-1) \epsilon_2+ a_\alpha -a_\beta~.
\eea
We then use identities involving the logarithm of BarnesÕ double gamma functions:\footnote{ We adopt the same convention for this function as in  \eg~ \cite{Nekrasov:2003rj, Alday:2009aq, Ito:2013kpa};  note that this is  different from that used in \eg~ \cite{Bonelli:2012ny}.}
\begin{equation}
\gamma_{\epsilon_1,\epsilon_2}(x)=\log \Gamma_2(x+\epsilon_+|\epsilon_1,\epsilon_2).
\end{equation}
where for $\epsilon_1>0$, $\epsilon_2<0$, we have an infinite product formula
 \begin{equation}
\Gamma_2(x|\epsilon_1,\epsilon_2) \propto \prod_{m,n \ge 1}
\left(x+(m-1)\epsilon_1- n\epsilon_2\right)^{+1}.\label{infiniteproduct2}
\end{equation} 
Thus we arrive at
\bea
Z^{\text{1-loop}, \BC^2}_{\text{vector},U(N)} (\vec a) =  \prod_{1\leq \alpha, \beta \leq N} \exp \Big[-\gamma_{\epsilon_1,\epsilon_2} (a_\alpha -a_\beta - \epsilon_1)-\gamma_{\epsilon_1,\epsilon_2} (a_\alpha -a_\beta - \epsilon_2) \Big]~.\label{oneloop-vector}
\eea
This is in agreement with (B.21) of \cite{Alday:2009aq}.
Similarly, for $SU(N)$ gauge group,
\bea
Z^{\text{1-loop}, \BC^2}_{\text{vector},SU(N)} (\vec a) &=  \exp \Big[\gamma_{\epsilon_1,\epsilon_2} (- \epsilon_1)+\gamma_{\epsilon_1,\epsilon_2} (- \epsilon_2)\Big]  \times\nn \\
& \quad \prod_{1\leq \alpha, \beta \leq N} \exp \Big[-\gamma_{\epsilon_1,\epsilon_2} (a_\alpha -a_\beta- \epsilon_1)-\gamma_{\epsilon_1,\epsilon_2} (a_\alpha -a_\beta- \epsilon_2) \Big]~.\label{oneloop-vector}
\eea

\subsubsection*{The orbifold case} 
Let us now turn to the orbifold case, focusing for the sake of concreteness on the $\mathbb{C}^2/\mathbb{Z}_2$ orbifold. Furthermore, for simplicity, we will consider the case where the orbifold acts trivially on the gauge fugacities, that is, $\vec r=0$.\footnote{Note that this is equivalent to consider the $U(1)$ case, as then the orbifold cannot act on the gauge fugacity --nevertheless absent for the gauge field, being in the adjoint representation--.}  Applying (\ref{prescription}) to project to orbifold-invariant states we find

\begin{equation}
{\rm ind}[D_{\rm SD}(\mathbb{C}^2/\mathbb{Z}_2)](t_1, t_2)=\frac{(1+t_1\,t_2)^2}{(1-t_1^2)\,(1-t_2^2)}\,\chi_{\rm \mathbf{Adj}}(\vec{z})\, .
\end{equation}
 
 Removing the Haar measure part the vector multiplet single-particle index becomes
\bea
\label{vectorORB}
f_{\vect}^{\rm \mathbb{C}^2/\mathbb{Z}_2} (t_1, t_2, \vec z; \vec r=0) 
&= (1-{\rm ind}[D_{\rm SD}(\BC^2/\BZ_2)])\,\chi_{\rm\mathbf{\adj}}(\vec z) \nn \\
&=-\frac{(t_1+t_2)^2}{(1-t_1^2)\,(1-t_2^2)}\,\chi_{\adj} (\vec z)~.
\eea

In general, for $\BC^2/\BZ_n$ with $\vec r=0$, we obtain
\bea
f_{\vect}^{\rm \mathbb{C}^2/\mathbb{Z}_n} (\vec t, \vec z; \vec r=0)= -\frac{t_1^n+t_2^n+2t_1 t_2(1-t_1^{n-1} t_2^{n-1})(1-t_1t_2)^{-1}}{\left(1-t_1^n\right) \left(1-t_2^n\right)} \chi_{\adj} (\vec z)~. \label{genorbnvecr0}
\eea

So far we have focused on trivial actions of the orbifold on the gauge fugacities. 
Computing the vector multiplet index contribution for general action $\vec r$ is straightforward albeit a bit more tedious. We discuss this issue for the large orbifold limit in Section \ref{largeorbifold}.

\subsubsection{Hypermultiplet contribution} \label{sec:hypercontr}
In the case of the hypermultiplet the relevant complex is the Dirac complex. In the unorbifolded case, borrowing the result for the Dirac complex index from  \cite{Kim:2012gu}, we have (we strip off gauge dependence)
\begin{equation}
\label{Dirac}
{\rm ind}[D_{\rm Dirac}](\mathbb{C}^2)(t_1,t_2)=\frac{\sqrt{t_1\,t_2}}{(1-t_1)\,(1-t_2)}
\end{equation}
For the matter multiplet the ${\rm PE}$ of the equivariant index of the complex is directly the contribution to the single-particle index. Thus, the contribution of the hypermultiplet in the representation ${\mathbf R}$ to the parition function is then given by
\bea
H^{\text{1-loop}, \BC^2}_{\text{matter}, {\mathbf R}} (t_1, t_2,\vec z,u) =  {\PE}[f^{\BC^2}_{{\bf R}}(t_1,t_2, \vec z) u^{+1/-1}]
\eea
with $u$ a flavour fugacity, the power of $u$ is $-1$ if ${\bf R}$ is the fundamental representation and $+1$ for other representations, and
\bea\label{fC2matter}
f^{\BC^2}_{{\bf R}} (t_1, t_2, \vec z) &={\rm ind}[D_{\rm Dirac}](\mathbb{C}^2)\, \chi_{\mathbf{\rm R}} (\vec z) \nn \\
&= \frac{\sqrt{t_1\,t_2}}{(1-t_1)\,(1-t_2)}\, \chi_{\mathbf{\rm R}} (\vec z)~.
\eea
Writting this with $t_1=x y, t_2=xy^{-1}$, one recovers the contribution in \cite{Kim:2012gu}.

\subsubsection*{Reduction on $S^1$} 
Using the variables as in \eref{Nekvars} and taking
\bea
u = e^{-\beta \mu}~
\eea
where $\mu$ is the mass parameter, we see that the one-loop part of the matter contributions in the $4d$ partition function are given as follows, upon the limit $\beta \rightarrow 0$: 
 \begin{align}
Z_\text{fund}^\text{1-loop}(\vec a,\mu) &=
\prod_{\alpha} \exp \left[\gamma_{\epsilon_1,\epsilon_2} (a_\alpha-\mu-\epsilon_+/2) \right],
\label{oneloop-hyper}\\
Z_\text{antifund}^\text{1-loop}(\vec a,\mu) &=
\prod_{\alpha} \exp \left[\gamma_{\epsilon_1,\epsilon_2} (-a_\alpha+\mu-\epsilon_+/2) \right],\\
Z_\text{bifund}^\text{1-loop}(\vec a,\vec b,m) &=
\prod_{\alpha,\beta} \exp \left[\gamma_{\epsilon_1,\epsilon_2} (a_\alpha-b_\beta-m-\epsilon_+/2) \right], \\
Z_\text{adjoint, $U(N)$}^\text{1-loop}(\vec a,m) &=
\prod_{\alpha,\beta=1}^N \exp \left[\gamma_{\epsilon_1,\epsilon_2} (a_\alpha-a_\beta-m-\epsilon_+/2) \right]~.
\end{align}
These formulae matches the expressions in (B.22)-(B.24) of \cite{Alday:2009aq}, with all mass parameters $\mu$ and $m$ shifted by $\epsilon_+/2$ with respect to those in \cite{Alday:2009aq}.

\subsubsection*{The orbifold case} 
As for the case of the vector multiplet, let us, for concreteness, concentrate on the case of $\mathbb{C}^2/\mathbb{Z}_2$ with trivial orbifold action $\vec r=0$ on the gauge fugacities. Following the general recipe (\ref{prescription}) we find

\begin{equation}
\label{indORB}
{\rm ind}[D_{\rm Dirac}(\mathbb{C}^2/\mathbb{Z}_2)]=\frac{\sqrt{t_1\,t_2}\,(1+t_1\,t_2)}{(1-t_1^2)\,(1-t_2^2)}\,\chi_{\rm \mathbf{R}}(\vec{z})
\end{equation}

Thus the relevant $f_{\rm matter}$ reads
\begin{equation}
\label{matterORB}
f_{{\bf R}}^{\rm \mathbb{C}^2/\mathbb{Z}_2} (\vec t, \vec z, u; \vec r=0)= \frac{\sqrt{t_1\,t_2}\,(1+t_1\,t_2)}{(1-t_1^2)\,(1-t_2^2)} \chi_{\mathbf{\rm R}} (\vec z)~.
\end{equation}
Again, for more complicated actions $\vec r$ of the orbifold on the gauge fugacities the corresponding expression for the matter contribution will be slightly more involved.  We discuss this issue for the large orbifold limit in Section \ref{largeorbifold}.

\subsection{Instanton contribution}
As described above, the 5d index contains a contribution from instantonic operators. Concentrating first on the unorbifolded case, such contribution factorizes into the contribution of instantons localized around the south pole and anti-instantons localized around the north pole of the $S^4$ \cite{Kim:2012gu}. Let us denote the instanton partition function for instantons around the south pole by $\mathcal{I}_{{\rm inst}}^{\rm S}$. Denoting by $q$ the instanton current fugacity, such function can be expanded as
\begin{equation}
\mathcal{I}_{\rm inst}^{\rm S}=\sum_{k=0}^{\infty}H^{\mathbb{C}^2}_{k}\,q^k\, , \label{suminstC2}
\end{equation}
so that $H^{\mathbb{C}^2}_{ k}$ is the $k$-instanton partition function (of course, $H^{\mathbb{C}^2}_{0}=1$).  Note that the explicit expressions of $H^{\mathbb{C}^2}_{ k}$ are very complicated\footnote{Explicit expressions for $k=2$ instantons on $\BC^2$ with various simple groups can be found in \cite{Hanany:2012dm}.} for larger values $k$, a closed form of the summation \eref{suminstC2} is not known.

On the other hand, the instanton index for anti-instantons localized around the north pole can be easily obtained  \cite{Kim:2012gu} as $\mathcal{I}_{\rm inst}^{\rm N}(q)=\mathcal{I}_{\rm inst}^{\rm S}(q^{-1})$. Then, the whole instanton contribution to the index is just $\mathcal{I}_{\rm inst}=\mathcal{I}_{\rm inst}^S\,\mathcal{I}_{\rm inst}^{\rm N}$. It is then clear that the quantities of interest are the $k$-instanton partition functions $H_k^{\mathbb{C}^2}$. 

Before turning to the orbifold case, let us briefly review the computation of the instanton contributions on $\mathbb{C}^2$.

\subsubsection{Instantons in gauge theories on $\BC^2$}

For pure gauge theories, these can be computed as the appropriately covariantized Hilbert series of the $k$-instanton moduli space \cite{Rodriguez-Gomez:2013dpa}. More generically, the contribution associated to the $k$-instanton for a generic theory can be computed using localization, which fixes the gauge field configuration such that the instantons are located at the origin. Such fixed instantons are labelled by an $N$-tuple of Young diagrams, denoted by $\vec Y = (Y_1, Y_2, \ldots, Y_N)$.  We refer to each element by $Y_\alpha$, with $\alpha=1, \ldots, N$, and we allow the cases in which there exist empty diagrams. The instanton number $k$ is given by the total number of boxes
\bea
k = |\vec Y| := \sum_{\alpha=1}^N | Y_\alpha|~.
\eea

For a given box $s$ at the $a$-th row and $b$-th column of a given Young diagram $Y$, one can define $a_Y(s)$ and $l_Y(s)$, known as the arm length and the leg length, as follows:
\bea
a_Y(s) = \lambda_a- b~, \qquad l_Y(s) = \lambda'_b - a~,
\eea
where $\lambda'_b$ corresponds to the transpose diagram of $Y$, namely $Y^T = (\lambda'_1 \geq \lambda'_2 \geq \ldots)$.  

The contribution from the vector multiplet is given by
\bea
&H^{\BC^2}_{\text{vector}} (t_1, t_2, \vec z; \vec Y) \nn \\
&=   \PE \left[ \sum_{\alpha, \beta=1}^N \sum_{s \in Y_\alpha } \left(\frac{z_\alpha}{z_\beta} t_1^{-l_{Y_\beta}(s)}t_2^{1+a_{Y_\alpha }(s)}+ \frac{z_\beta}{z_\alpha} t_1^{1+l_{Y_\beta }(s)}t_2^{-a_{Y_\alpha}(s)} \right) \right] ~, \label{eq:HSC2vec}
\eea
Note that if $s \in Y_\alpha$ but $s \notin Y_\beta$, then $l_{Y_\beta}(s)$ can be negative.
The contributions from the fundamental and antifundamental hypermultiplets are given by
\bea
& H^{\BC^2}_{\text{fund}} (t_1, t_2, \vec z,  u; \vec Y) =   \PE \left[ u^{-1}  \sum_{\alpha=1}^N z_\alpha \sum_{(a,b) \in Y_\alpha }  t_1^a t_2^b \right] ~, \label{eq:HSC2fund} \\
& H^{\BC^2}_{\text{antifund}} (t_1, t_2, \vec z, u; \vec Y) =\PE \left[ u \sum_{\alpha=1}^N z_\alpha  \sum_{(a,b) \in Y_\alpha }  t_1^{a-1} t_2^{b-1} \right] ~, \label{eq:HSC2antifund} \\
& H^{\BC^2}_{\text{adjoint}} (t_1, t_2, \vec z,u ; \vec Y) \nn \\
& \hspace{2cm} = \PE \left[ u\sum_{\alpha, \beta=1}^N \sum_{s \in Y_\alpha } \left(\frac{z_\alpha}{z_\beta} t_1^{-l_{Y_\beta}(s)}t_2^{1+a_{Y_\alpha }(s)}  + \frac{z_\beta}{z_\alpha} t_1^{1+l_{Y_\beta }(s)}t_2^{-a_{Y_\alpha}(s)} \right) \right]~.\label{eq:HSC2adjoint} 
\eea
where $u$ denotes the fugacity for the flavour symmetry.

The contribution from the instanton number $k$ is given by
\bea\label{eq:sumyoungC2}
H^{\BC^2}_{\text{inst}, k, U(N)} (\vec t, \vec z, \vec u)  = \sum_{\vec Y: |\vec Y| =k} \frac{H^{\BC^2}_{\text{vector}} (\vec t, \vec z; \vec Y) }{H^{\BC^2}_{\text{matter}} (\vec t, \vec z, \vec u; \vec Y)}~.
\eea 
where the summation runs over all possible $N$-tuples of the Young diagrams whose total number of boxes equal to the instanton number $k$.

\subsubsection*{Example: one instanton contribution to $SU(2)$ with $4$ flavours}

In general, for an $SU(N)$ theory with $N_f=2\,N$, the $k$-instanton contribution is

\bea
H^{\BC^2}_{\text{inst}, k} (t_1, t_2, \vec z, \vec u)  &= \sum_{\vec Y: |\vec Y| =k} \frac{H^{\BC^2}_{\text{vector}} (t_1,t_2, \vec z; \vec Y)}{\prod_{i=1}^N H^{\BC^2}_{\text{antifund}} (t_1,t_2, \vec z, u_i; \vec Y) \prod_{j=N+1}^{2N} H^{\BC^2}_{\text{fund}} (t_1,t_2, \vec z, u_j; \vec Y)} ~.
\eea

Focusing on the $N=2$ case, the ordered pairs of Young diagrams that contribute to the partition function are
\bea
 (\Box, \emptyset)~, \quad  (\emptyset, \Box)~.
\eea
Here are the contributions for each part:
\bea
H^{\BC^2}_{\text{vector}} (t_1,t_2, \vec z; \vec (\Box, \emptyset)) &=  \PE \left[t_1+t_2+\frac{t_1 t_2 z_1}{z_2}+\frac{z_2}{z_1} \right] \\
H^{\BC^2}_{\text{fund}} (t_1,t_2, \vec z,u ; \vec (\Box, \emptyset)) &=\PE \left[ \frac{t_1 t_2 z_1}{u} \right] \\
H^{\BC^2}_{\text{antifund}} (t_1,t_2, \vec z,u ; \vec (\Box, \emptyset)) &= \PE \left[ uz_1 \right]~.
\eea
The contribution from $(\emptyset, \Box)$ can be obtained from above by exchanging $z_1$ and $z_2$.

Therefore, the one-instanton contribution is given by
\bea
H^{\BC^2}_{\text{inst}, k=1} (t_1, t_2, \vec z, \vec u) 
&= \PE \left[t_1+t_2+\frac{t_1 t_2 z_1}{z_2}+\frac{z_2}{z_1} -  \frac{t_1 t_2 z_1}{u} - uz_1  \right] + (z_1 \leftrightarrow z_2) \label{HC2instk1}\\
&= \frac{\prod_{i=1}^2(1-u_i z_1^{-1}) \prod_{j=3}^4 (1-t_1 t_2 z_1 u_j^{-1})}{(1-t_1)(1-t_2)(1-t_1 t_2 z_1 z_2^{-1})(1- z_2 z_1^{-1})} + (z_1 \leftrightarrow z_2)~.
\eea
The $4d$ limit of this contribution is
\bea
&Z^{\BC^2}_{\text{inst}, k=1} (\epsilon_1, \epsilon_2, \vec a, \vec \mu)\nn \\
&= \lim_{\beta \rightarrow 0} H^{\BC^2}_{\text{inst}, k=1} (e^{-\beta \epsilon_1}, e^{-\beta \epsilon_2}, e^{-\beta \vec a}, e^{-\beta \mu}) \nn \\
&=\frac{\left(a_1+\mu _1\right) \left(a_1+\mu _2\right) \left(a_1+\epsilon _1+\epsilon _2-\mu _3\right) \left(a_1+\epsilon _1+\epsilon _2-\mu _4\right)}{\left(-a_1+a_2\right) \epsilon _1 \epsilon _2 \left(a_1-a_2+\epsilon _1+\epsilon _2\right)} + (a_1 \leftrightarrow a_2)~.
\eea
This is in agreement with (A.7) of \cite{Ito:2013kpa}.

\subsubsection{Instantons in gauge theories on $\BC^2/\BZ_n$}
In the orbifold case we should compute the instanton partition functions on $\mathbb{C}^2/\mathbb{Z}_n$. For a given $U(N)$ gauge theory on $\BC^2/\BZ_n$, upon choosing the holonomy $\vec r$, the instanton partition function with Kronheimer-Nakajima vector $\vec k$ for such a theory, denoted by $H_{\vec{k},\,\vec{r}}^{\mathbb{C}^2/\mathbb{Z}_n}$, can be directly obtained from the case of $\BC^2$.  One simply needs to apply the following implementations for the orbifold projection:
\ben
\item In Eq. \eref{eq:sumyoungC2}, the summation runs over a certain set $\CR( \vec k, \vec r)$ of tuples of Young diagram defined as follows.  Given $\vec r$ and $\vec k$, $\CR( \vec k, \vec r)$ is a set of $N$-tuples of Young diagrams such that all of the following conditions are satisfied:
\ben
\item The total number of boxes in $\vec Y$ is given by $|\vec Y| := \sum_{\alpha=1}^N Y_\alpha = \sum_{i=1}^n k_i$.
\item Upon assigning the numbers $r_\alpha+a-b~(\mathrm{mod}~n)$ to all $(a,b)$ boxes of every non-trivial Young diagram $Y_\alpha \neq \emptyset$ for all $\alpha=1, \ldots, N$, there must be precisely $k_j$ boxes in total that are labelled by the number $j~(\mathrm{mod}~n)$ for all $j=1, \ldots, n$.
\een
\item Only the terms inside the $\PE$s in Eqs. \eref{eq:HSC2vec}, \eref{eq:HSC2fund}, \eref{eq:HSC2antifund} and \eref{eq:HSC2adjoint} that are invariant under the actions \eref{actiont} and \eref{actionz} are kept; the other terms are thrown away.
\ben
\item For the vector multiplet and adjoint hypermultiplet contributions, the summation over $\alpha, \beta$ and $s$ in \eref{eq:HSC2vec} and \eref{eq:HSC2adjoint} are restricted to those satisfying
\bea
-r_\alpha + r_\beta + l_{Y_\beta}(s)+ a_{Y_\alpha}(s)+1=0 \quad ({\rm mod}~n)~.
\eea
\item For the fundamental and antifundamental hypermultiplet contributions, the summation over $\alpha$ and $(a,b)$ in \eref{eq:HSC2fund} and \eref{eq:HSC2antifund} are restricted to those satisfying
\bea
r_\alpha+a -b =0 \quad ({\rm mod}~n)~.
\eea
\een
Explicitly,
{\small
\bea
&H^{\BC^2/\BZ_n}_{\text{vector}} (\vec t, \vec z; \vec Y; \vec r) =   \PE \Bigg[ \sum_{\alpha, \beta=1}^N \sum_{s \in Y_\alpha } \left(\frac{z_\alpha}{z_\beta} t_1^{-l_{Y_\beta}(s)}t_2^{1+a_{Y_\alpha }(s)}+ \frac{z_\beta}{z_\alpha} t_1^{1+l_{Y_\beta }(s)}t_2^{-a_{Y_\alpha}(s)} \right)  \times \nn \\
& \hspace{6cm} \delta_{-r_\alpha + r_\beta + l_{Y_\beta}(s)+ a_{Y_\alpha}(s)+1 \;({\rm mod}~n),0} \Bigg] ~,  \\ 
& H^{\BC^2/\BZ_n}_{\text{fund}} (\vec t, \vec z,  u; \vec Y; \vec r) =   \PE \Bigg[ u^{-1}  \sum_{\alpha=1}^N z_\alpha \sum_{(a,b) \in Y_\alpha }  t_1^a t_2^b~ \delta_{r_\alpha+a-b \;({\rm mod}~n),0}\Bigg] ~,  \\
& H^{\BC^2/\BZ_n}_{\text{antifund}} (t_1, t_2, \vec z, u; \vec Y; \vec r) =\PE \left[ u \sum_{\alpha=1}^N z_\alpha  \sum_{(a,b) \in Y_\alpha }  t_1^{a-1} t_2^{b-1} \delta_{r_\alpha+a-b \;({\rm mod}~n),0} \right] ~, \\
& H^{\BC^2/\BZ_n}_{\text{adjoint}} (t_1, t_2, \vec z,u ; \vec Y; \vec r)  = \PE \Bigg[ u\sum_{\alpha, \beta=1}^N \sum_{s \in Y_\alpha } \left(\frac{z_\alpha}{z_\beta} t_1^{-l_{Y_\beta}(s)}t_2^{1+a_{Y_\alpha }(s)}  + \frac{z_\beta}{z_\alpha} t_1^{1+l_{Y_\beta }(s)}t_2^{-a_{Y_\alpha}(s)} \right)  \times \nn \\
& \hspace{6cm} \delta_{-r_\alpha + r_\beta + l_{Y_\beta}(s)+ a_{Y_\alpha}(s)+1 \;({\rm mod}~n),0} \Bigg] ~.
\eea}
where $u$ denotes the fugacity for the flavour symmetry.  It is chosen in such a way that the $4d$ limit is in accordance with the convention of \cite{Ito:2013kpa}; see \eg~\eref{NekU24flvk11r00} below.  Note that in Section \ref{monopolebubbling}, we make a redefinition of $u$ so that the results are in agreement with those in \cite{Ito:2011ea, Gang:2012yr}.

\item The $5d$ instanton partition function (or Hilbert series) for $U(N)$ gauge theory on $\BC^2/\BZ_2$ is given by
\bea
H^{\BC^2/\BZ_n}_{\vec k, \vec r} (\vec t, \vec z, \vec u)  = \sum_{\vec Y \in \CR( \vec k, \vec r)} \frac{H^{\BC^2/\BZ_n}_{\text{vector}} (\vec t, \vec z; \vec Y) }{H^{\BC^2/\BZ_n}_{\text{matter}} (\vec t, \vec z, \vec u; \vec Y)}~.
\eea 
\een

\subsubsection*{Reduction on $S^1$} 

Upon reduction along the ``time" $S^1$, the instanton contribution, both in the orbifold and orbifolded cases, does go over to the known instanton partition function for 4d theories on an orbifold. We refere to \cite{DHMRS2013} and references therein for explicit expressions. Note that in \cite{DHMRS2013} the quantity computed is the Hilbert series of the instanton moduli space. Nevertheless the Nekrasov instanton partition function directly follows up to multiplication by the suitable factor of $x$ \cite{Rodriguez-Gomez:2013dpa}. Since this factor in the 4d limit simply becomes 1, the reductions in \cite{DHMRS2013} and the described matchings with the known expressions in the literature for instantons on ALE spaces can be borrowed in the case at hand to conclude that indeed the non-perturbative part of the 5d index, reduced along the time $S^1$, becomes the non-perturbative contribution to the partition function of the gauge theory on the ALE space.

Thus, in view of the reduction of both the perturbative and non-perturbative contributions to the 5d index on $S^1\times S^4/\mathbb{Z}_n$, all in all we find that the reduction along the ``temporal" $S^1$ does indeed recover the partition function of the 4d version of the theory on the ALE space.

\vspace{1cm}

A number of examples for instantons in $\CN=2$ $U(N)$ and $SU(N)$ pure gauge theory are presented in \cite{DHMRS2013}.  We shall not repeat them here; however, in the following, we present some examples for gauge theories with matter.

\subsubsection{Example: $SU(2)$ theory with one hypermultiplet on $\BC^2/\BZ_n$}
\subsubsection* {$1/2$ pure instantons on $\BC^2/\BZ_2$ with $\vec r =(1,1)$: $\vec k=(1,0)$}
The set $\CR(\vec k =(1,0), \vec r=(1,1))$ contains the following elements:
\bea
\vec Y_1= (\emptyset, \tiny \yng(1)),\qquad  \vec Y_2 = (\tiny \yng(1), \emptyset) ~. \label{Yngr11}
\eea

The contributions of the vector multiplet and the adjoint hypermultiplet are
\bea
H^{\BC^2/\BZ_2}_{\text{vector}} (t_1,t_2, \vec z; \vec Y_2) &=  \PE \left[\frac{t_1 t_2 z_1}{z_2}+\frac{z_2}{z_1}\right]~, \label{vecr11} \\
H^{\BC^2/\BZ_2}_{\text{adjoint}} (t_1,t_2, \vec z,u ; \vec Y_2) &=\PE \left[ \frac{u t_1 t_2 z_1}{z_2}+\frac{u z_2}{z_1} \right]~.
\eea
For $\vec Y_1$, one only needs to exchange $z_1$ and $z_2$.

Hence, the instanton partition function is given by
\bea \label{5dN4SYM1011}
H^{\BC^2/\BZ_2}_{\text{inst}; \vec k=(1,0), \vec r=(1,1)} (t_1,t_2,\vec z, \vec u) =  \frac{\left(1-\frac{u z_1}{z_2}\right) \left(1-\frac{u t_1 t_2 z_2}{z_1}\right)}{\left(1-\frac{z_1}{z_2}\right) \left(1-\frac{t_1 t_2 z_2}{z_1}\right)}+ (z_1 \leftrightarrow z_2)~.
\eea 
The $4d$ limit of this expression is
\bea\label{4dN4SYM1011}
&Z^{\BC^2/\BZ_2}_{\text{inst};  \vec k=(1,0), \vec r=(1,1)} (\epsilon_1, \epsilon_2, \vec a, \vec \mu) \nn \\
&= \lim_{\beta \rightarrow 0} \beta^{-2} H^{\BC^2/\BZ_2}_{\text{inst};  \vec k=(1,0), \vec r=(1,1)} (e^{-\beta \epsilon_1}, e^{-\beta \epsilon_2}, e^{-\beta \vec a}, e^{-\beta \mu}) \nn \\
&= \frac{\left(a_1-a_2+\mu\right) \left(-a_1+a_2+\epsilon _1+\epsilon _2+\mu\right)}{\left(a_1-a_2\right) \left(-a_1+a_2+\epsilon _1+\epsilon _2\right)} + (a_1 \leftrightarrow a_2) \nn \\
&= -\frac{2 \left[\left(a_1-a_2\right){}^2-\left(\epsilon _1+\epsilon _2\right){}^2-\left(\epsilon _1+\epsilon _2\right) \mu-\mu^2\right]}{\left(a_1-a_2+\epsilon _1+\epsilon _2\right) \left(-a_1+a_2+\epsilon _1+\epsilon _2\right)}~.
\eea
We shall make use of Eqs. \eref{5dN4SYM1011} and \eref{4dN4SYM1011} later in Section \ref{sec:N2starU220}.

\subsubsection*{$2/3$ pure instantons on $\BC^2/\BZ_3$ with $\vec r =(1,2)$: $\vec k=(1,1,0)$} \label{sec:N2starU212}
The set $\CR(\vec k =(1,1,0), \vec r=(1,2))$ contains the following elements:
\bea
\vec Y_1 = ( \emptyset, (2)) , \quad \vec Y_2 = ((1), (1)), \quad \vec Y_3 = ((1, 1), \emptyset) ~.
\eea
The contributions of the vector multiplet and the adjoint hypermultiplet are
\bea
H^{\BC^2/\BZ_3}_{\text{vector}} (\vec t, \vec z; \vec Y_i) &=  \left( \PE \left[ \frac{z_1}{t_2 z_2}+\frac{t_1 t_2^2 z_2}{z_1} \right], \PE \left[ \frac{t_1 z_1}{z_2}+\frac{t_2 z_2}{z_1} \right] , \PE \left[ \frac{t_1^2 t_2 z_1}{z_2}+\frac{z_2}{t_1 z_1} \right] \right)~, \nn \\
H^{\BC^2/\BZ_3}_{\text{adjoint}} (\vec t, \vec z,u ; \vec Y_i) &=\left( \PE \left[u \left( \frac{z_1}{t_2 z_2}+\frac{t_1 t_2^2 z_2}{z_1} \right) \right], \PE \left[u \left( \frac{t_1 z_1}{z_2}+\frac{t_2 z_2}{z_1} \right) \right] , \right. \nn \\
& \qquad \left. \PE \left[ u \left( \frac{t_1^2 t_2 z_1}{z_2}+\frac{z_2}{t_1 z_1} \right) \right] \right)~.
\eea
Hence, the instanton partition function is given by
\bea 
H^{\BC^2/\BZ_3}_{\text{inst}; \vec k=(1,1,0), \vec r=(1,2)} (\vec t,\vec z, u) &=\frac{\left(1-\frac{u t_1^2 t_2 z_1}{z_2}\right) \left(1-\frac{u z_2}{t_1 z_1}\right)}{\left(1-\frac{t_1^2 t_2 z_1}{z_2}\right) \left(1-\frac{z_2}{t_1 z_1}\right)}+\frac{\left(1-\frac{u t_1 z_1}{z_2}\right) \left(1-\frac{u t_2 z_2}{z_1}\right)}{\left(1-\frac{t_1 z_1}{z_2}\right) \left(1-\frac{t_2 z_2}{z_1}\right)} \nn \\
& \quad +\frac{\left(1-\frac{u z_1}{t_2 z_2}\right) \left(1-\frac{u t_1 t_2^2 z_2}{z_1}\right)}{\left(1-\frac{z_1}{t_2 z_2}\right) \left(1-\frac{t_1 t_2^2 z_2}{z_1}\right)}~. \label{5dN2starr12}
\eea
We shall make use of \eref{5dN2starr12} later in Section \ref{sec:N2starU230}.

\subsubsection{Example: $U(2)$ gauge theory with $4$ flavours on $\BC^2/\BZ_2$}
\subsubsection*{Instantons on $\BC^2/\BZ_2$ with $\vec k = (1,2)$ and $\vec r= (0,0)$}
The tuples of Young diagrams that contribute to the partition functions are 
\bea
\vec Y_1 =(\emptyset, \tiny \yng(1,2)), \qquad  \vec Y_2= (\tiny \yng(1,2), \emptyset)~.
\eea
For example, the contribution of $\vec Y_2$ to the vector multiplet part is
\bea
H^{\BC^2}_{\text{vector}} (t_1,t_2, \vec z; \vec Y_2) &= \PE \left[ 2 t_1+\frac{t_1^2}{t_2}+2 t_2+\frac{t_2^2}{t_1}+\frac{t_1 t_2 z_1}{z_2}+\frac{t_1^2 t_2 z_1}{z_2}+\frac{t_1 t_2^2 z_1}{z_2}+\frac{z_2}{z_1} \right. \nn \\
& \qquad \left.+\frac{z_2}{t_1 z_1}+\frac{z_2}{t_2 z_1} \right]~.
\eea
After keeping only terms in the $\PE$ that are invariant under \eref{actiont} and \eref{actionz}, the contribution of the vector multiplet is
\bea
H^{\BC^2/\BZ_2}_{\text{vector}} (t_1,t_2, \vec z; \vec Y_2) &=  \PE \left[\frac{t_1 t_2 z_1}{z_2}+\frac{z_2}{z_1} \right]~.
\eea
Similarly, the contributions of the hypermultiplets are
\bea
H^{\BC^2/\BZ_2}_{\text{fund}} (t_1,t_2, \vec z,u ; \vec Y_2) &=\PE \left[ \frac{t_1 t_2 z_1}{u} \right]~, \\
H^{\BC^2/\BZ_2}_{\text{antifund}} (t_1,t_2, \vec z,u ; \vec Y_2) &= \PE \left[ uz_1 \right]~.
\eea
For $\vec Y_1$, one only needs to exchange $z_1$ and $z_2$.

Thus, the one-instanton contribution is given by
\bea
& H^{\BC^2/\BZ_2}_{\text{inst}; \vec k=(1,2), \vec r=(0,0)}( t_1,t_2,\vec z, \vec u)   \nn \\
&= \PE \left[ \frac{t_1 t_2 z_1}{z_2}+\frac{z_2}{z_1} - \sum_{j=3}^4 \frac{t_1 t_2 z_1}{u_j} - \sum_{i=1}^2 u_iz_1 \right] +(z_1 \leftrightarrow z_2)\nn \\
&= \frac{\left(1-u_1 z_1\right) \left(1-u_2 z_1\right) \left(1-\frac{t_1 t_2 z_1}{u_3}\right) \left(1-\frac{t_1 t_2 z_1}{u_4}\right)}{\left(1-\frac{t_1 t_2 z_1}{z_2}\right) \left(1-\frac{z_2}{z_1}\right)} +(z_1 \leftrightarrow z_2)~.
\eea
The $4d$ limit of this contribution is
\bea
&Z^{\BC^2/\BZ_2}_{\text{inst}; \vec k, \vec r} (\epsilon_1, \epsilon_2, \vec a, \vec \mu) \nn \\
&= \lim_{\beta \rightarrow 0} \beta^{-2} H^{\BC^2/\BZ_2}_{\text{inst}, \vec k=(1,1)} (e^{-\beta \epsilon_1}, e^{-\beta \epsilon_2}, e^{-\beta \vec a}, e^{-\beta \mu}) \nn \\
&= \frac{\left(a_1+\mu _1\right) \left(a_1+\mu _2\right) \left(a_1+\epsilon _1+\epsilon _2-\mu _3\right) \left(a_1+\epsilon _1+\epsilon _2-\mu _4\right) }{\left(-a_1+a_2\right) \left(a_1-a_2+\epsilon _1+\epsilon _2\right)}+ (a_1 \leftrightarrow a_2)~. \label{NekU24flvk11r00}
\eea
This is in agreement with (A.8) of \cite{Ito:2013kpa}.

\subsubsection*{Instantons on $\BC^2/\BZ_2$ with $\vec k=(0,1)$ and $\vec r =(0,0)$}
The ordered pairs $\vec Y_a$ (with $a=1,2$) of Young diagrams that contribute to the Hilbert series are given by \eref{Yngr11}.  The contribution from the vector multiplet is given by \eref{vecr11}. The contribution from the fundamental hypermultiplet is
\bea
H^{\BC^2/\BZ_2}_{\text{fund}} (\vec t, \vec z,u ; \vec Y_a) &=   \left( \PE\left[ \frac{t_1 t_2 z_2}{u} \right] , \PE \left[ \frac{t_1 t_2 z_1}{u} \right] \right )~, \qquad a=1,2.
\eea
The contribution from the anti-fundamental hypermultiplet is
\bea
H^{\BC^2/\BZ_2}_{\text{antifund}} (\vec t, \vec z,u ; \vec Y_a) &= \left( \PE[ u z_2 ], \PE[ u z_1 ]\right)
\eea
The Hilbert series is given by
\bea \label{U24flvk10r11}
&H^{\BC^2/\BZ_2}_{\text{inst}; \vec k=(0,1), \vec r=(0,0)}( \vec t,\vec z, \vec u)  \nn \\
&= \sum_{a=1}^2 \frac{H^{\BC^2/\BZ_2}_{\text{vector}} (\vec t, \vec z; \vec Y_a) }{\prod_{i=1}^2 H^{\BC^2/\BZ_2}_{\text{antifund}} (\vec t, \vec z, u_i; \vec Y_a)\prod_{j =3}^4 H^{\BC^2/\BZ_2}_{\text{fund}} (\vec t, \vec z, u_j; \vec Y_a)} \nn \\
&=  \frac{\left(1-u_1 z_1\right) \left(1-u_2 z_1\right) \left(1-\frac{t_1 t_2 z_1}{u_3}\right) \left(1-\frac{t_1 t_2 z_1}{u_4}\right)}{\left(1-\frac{t_1 t_2 z_1}{z_2}\right) \left(1-\frac{z_2}{z_1}\right)} \nn \\
& \quad +\frac{\left(1-u_1 z_2\right) \left(1-u_2 z_2\right) \left(1-\frac{t_1 t_2 z_2}{u_3}\right) \left(1-\frac{t_1 t_2 z_2}{u_4}\right)}{\left(1-\frac{z_1}{z_2}\right) \left(1-\frac{t_1 t_2 z_2}{z_1}\right)}~.
\eea
We shall make use of this result later in Section \ref{sec:U24flvB20}.

\section{Large orbifold limit and relation to the $4d$ 't Hooft line index}\label{largeorbifold}

Given that our theories are placed on a $\mathbb{Z}_n$ orbifold background, it is natural to ask about the large orbifold limit.  Recall that we are considering theories on $S^1\times S^4/\mathbb{Z}_n$. Let us look more closely to the $S^4/\mathbb{Z}_n$ metric

\begin{equation}
ds^2_{S^4/\mathbb{Z}_n}=d\alpha^2+\frac{\sin\alpha^2}{4}\,(d\psi-\cos\theta\,d\phi)^2+\frac{\sin\alpha^2}{4}\,\Big( d\theta^2+\sin^2\theta\,d\phi^2\Big)
\end{equation}
where $\psi\in [0,\,\frac{4\,\pi}{n}]$. Defining $\hat{\psi}=\frac{n}{2}\,\psi\in [0,\,2\,\pi]$ we have

\begin{equation}
ds^2_{S^4/\mathbb{Z}_n}=d\alpha^2+\frac{\sin\alpha^2}{n^2}\,(d\hat{\psi}-\frac{n}{2}\,\cos\theta\,d\phi)^2+\frac{\sin\alpha^2}{4}\,\Big( d\theta^2+\sin^2\theta\,d\phi^2\Big)
\end{equation}
Thus, in the large orbifold limit $n\rightarrow \infty$ we roughly find

\begin{equation}
ds^2_{S^4/\mathbb{Z}_n}\rightarrow d\alpha^2+\frac{\sin\alpha^2}{4}\,\Big( d\theta^2+\sin^2\theta\,d\phi^2\Big)
\end{equation}
thus obtaining a 3d space (albeit with two conical singularities at both poles $\alpha=0,\,\pi$). Hence, since the large orbifold limit amounts to a dimensional reduction, we expect to recover, in large $n$, results for indices of 4d gauge theories.

Note that, in contrast to the reduction along the ``temporal" $S^1$, since we are not reducing along the $S^1$ along which the supersymmetric boundary conditions are set, instead of reducing the index to a partition function, in this case we should expect to find a 4d index. This is in fact very similar to the lens space indices very recently discussed in \cite{Benini:2011nc, Razamat:2013opa}.

The large orbifold limit is to be implemented simultaneously on both the perturbative and non-perturbative contributions to the 5d index. Unfortunately, computing the non-perturbative part for a generic orbifold $\mathbb{C}^2/\mathbb{Z}_n$ is technically challenging, so let alone taking the large orbifold limit. Thus, we will concentrate on the perturbative part obtaining quite amusing results. In section \ref{conclusions} we will come back to this point and speculate on the properties of the non-perturbative contribution.

\subsection{Perturbative part} \label{sec:pertinforb}

Let us discuss the large orbifold limit of the perturbative contribution to the index. In this limit, the discrete $\BZ_n$ action becomes a continuous $U(1)$ action, whose fugacity is denoted by $w$.  The orbifold action on $t_1$, $t_2$ and $z_\alpha$ (with $\alpha=1, \ldots, N$) is therefore
\bea
t_1 \rightarrow w t_1~, \qquad t_2 \rightarrow w^{-1} t_2~, \qquad z_\alpha \rightarrow w^{r_\alpha} z_\alpha~.
\eea

\subsubsection{The case of $\vec r=0$ and the Schur index}
Let us first consider the case of an orbifold acting trivially on the gauge group fugacities, namely $\vec r =0$. Using \eref{SDindex} and \eref{Dirac}, we see that in the large orbifold limit the self-dual and Dirac complexes are
{\small
\bea
 {\rm ind}[D_{\rm SD}](\mathbb{C}^2/\mathbb{Z}_{\infty})&= \oint_{|w|=1} \frac{\ud q}{(2\pi i)w} {\rm ind}[D_{\rm SD}(\mathbb{C}^2)](w t_1, w^{-1} t_2) 
 =  \frac{1+t_1t_2}{1-t_1t_2}\\
{\rm ind}[D_{\rm Dirac}](\mathbb{C}^2/\mathbb{Z}_{\infty}) &=  \oint_{|w|=1} \frac{\ud q}{(2\pi i)w} {\rm ind}[D_{\rm Dirac}(\mathbb{C}^2)](w t_1, w^{-1} t_2) 
=\frac{\sqrt{t_1 t_2}}{(1-t_1)(1-t_2)}~.
\eea}
Therefore, the vector and matter multiplet contributions to the index are
\bea
f_{\rm vector}^{\mathbb{C}^2/\mathbb{Z}_{\infty}} (\vec t, \vec z; \vec r =0)&= \left( 1- {\rm ind}[D_{\rm SD}](\mathbb{C}^2/\mathbb{Z}_{\infty})\right)=  -\frac{2\,t_1t_2}{1-t_1t_2}\,\chi_{\rm \mathbf{\adj}}(\vec z) \\ 
f_{\rm matter}^{\mathbb{C}^2/\mathbb{Z}_{\infty}}(\vec t, \vec z; \vec r=0)&= \frac{\sqrt{t_1t_2}}{1-t_1t_2}\,\chi_{\rm \mathbf{R}} (\vec z)~,
\eea
where we used the same notation as in Section \ref{sec:hypercontr}.
Amusingly, this is the 4d Schur index as described in (4.14) of \cite{Gadde:2011uv}. \footnote{ The fugacity $\rho$ (4.14) of \cite{Gadde:2011uv} is identified to ours as $\rho\equiv x$.}

Note that this result is only valid for the case of an orbifold with trivial action on the gauge group. The cases of general $\vec r$ are more involved, as we shall discuss below. 

\subsubsection{General $\vec r$ and the 't Hooft line perturbative index}
Let us start with the simple case of $U(2)$. The generic action of the orbifold on the gauge fugacities is
\begin{equation}
(z_1,\,z_2)\rightarrow (\omega^{r_1}\,z_1,\,\omega^{r_2}\,z_2)\, .
\end{equation}
 We then have
\begin{itemize}
\item \textbf{Vector multiplet} \\
Starting with the $\mathbb{C}^2$ self-dual complex, including the $U(2)$ gauge character
\begin{equation}
{\rm ind}[D_{\rm SD}](\mathbb{C}^2)=\frac{1+t_1t_2}{1-t_1t_2}\,\Big((z_1+z_2)\,(z_1^{-1}+z_2^{-1})\Big)
\end{equation}
we find, in the large orbifold limit, the following index
\begin{equation}
\label{vectorU(2)}
{\rm ind}[D_{\rm SD}](\mathbb{C}^2/\mathbb{Z}_{\infty})=\frac{1+t_1t_2}{1-t_1t_2}\,(t_1^{|r_1-r_2|}\,z_1\,z_2^{-1}+2+t_2^{|r_2-r_1|}\,z_2\,z_1^{-1})
\end{equation}
Had we considered the $SU(2)$ case, we would have found 
\begin{equation}
\label{vectorSU(2)}
{\rm ind}[D_{\rm SD}](\mathbb{C}^2/\mathbb{Z}_{\infty})=\frac{1+t_1t_2}{1-t_1t_2}\,(t_1^{|r_1-r_2|}\,z_1\,z_2^{-1}+1+t_2^{|r_2-r_1|}\,z_2\,z_1^{-1})
\end{equation}
Note that the difference between (\ref{vectorU(2)}) and (\ref{vectorSU(2)}) is a factor of the $U(1)$ self-dual complex index\footnote{For $U(1)$ the $\chi_{\rm \mathbf{\adj}}=1$, and hence there is no action of the orbifold on the gauge group. The contribution to the index is just $-\frac{2\,x^2}{1-x^2}$. This factor is indeed the difference between (\ref{vectorU(2)}) and (\ref{vectorSU(2)}).}, which is precisely what one would expect.

\item \textbf{Hypermultiplets in the (anti-) fundamental representation}

Starting now with the $\mathbb{C}^2$ Dirac complex, including the $U(2)$ character
\begin{equation}
{\rm ind}[D_{\rm Dirac}](\mathbb{C}^2)=\frac{\sqrt{t_1\,t_2}}{(1-t_2)\,(1-t_2)} \times 
\begin{cases} (z_1+z_2) & \text{fundamental}\\ (z_1^{-1}+z_2^{-1}) & \text{anti-fundamental}\end{cases}
\end{equation}
we now find, in the large orbifold limit, the following index
\begin{equation}
{\rm ind}[D_{\rm Dirac}](\mathbb{C}^2/\mathbb{Z}_{\infty})=\frac{\sqrt{t_1t_2}}{1-t_1t_2} \times
\begin{cases} t_2^{|r_1|} z_1+ t_1^{|r_2|}z_2  & \text{fundamental}\\ t_2^{|r_1|} z_1^{-1}+ t_1^{|r_2|}z_2^{-1} & \text{anti-fundamental}\end{cases}
\end{equation}
\item \textbf{Hypermultiplets in the adjoint representation}

Starting now with the $\mathbb{C}^2$ Dirac complex, including the $U(2)$ adjoint character
\begin{equation}
{\rm ind}[D_{\rm Dirac}](\mathbb{C}^2)=\frac{\sqrt{t_1\,t_2}}{(1-t_2)\,(1-t_2)}\,\Big(z_1+z_2)(z_1^{-1}+z_2^{-1})\Big)
\end{equation}
we now find, in the large orbifold limit, the following index

\begin{equation}
\label{hyperadjU(2)}
{\rm ind}[D_{\rm Dirac}](\mathbb{C}^2/\mathbb{Z}_{\infty})=\frac{\sqrt{t_1t_2}}{1-t_1t_2}\,(t_1^{|r_1-r_2|}\,z_1\,z_2^{-1}+2+t_2^{|r_2-r_2|}\,z_2\,z_1^{-1})
\end{equation}
Again, had we considered the $SU(2)$ case we would have found the same expression (\ref{hyperadjU(2)}) only that with a 1 instead of a 2.

\end{itemize}

However, while the Dirac complex is directly the single-letter contribution to the index, the self-dual complex contains information about the integration measure as described in section \ref{orbifoldindex}. In the case at hand, note that both the adjoint hypermultiplet and the vector multiplet contribute very similarly to the trivial monodromy case, only exchanging the adjoint character by the slightly more complicated factor $(t_1^{|r_1-r_2|}\,z_1\,z_2^{-1}+1+t_2^{|r_2-r_1|}\,z_2\,z_1^{-1})$. Thus, it is natural to extract in this case the factor ${\rm PE}[-(t_1^{|r_1-r_2|}\,z_1\,z_2^{-1}+1+t_2^{|r_2-r_1|}\,z_2\,z_1^{-1})]$ from the self-dual complex index to find the vector-multiplet contribution. Note that for the trivial action case $\vec{r}=0$ this recovers the adjoint character, and hence the result in \ref{orbifoldindex}. Thus, the appropriate generalized measure is, in this case, given by\footnote{The $\frac{1}{2}$ ensures the correct normalization.}
\bea
 \int [ \ud \vec z]_{\vec r} =\frac{1}{2}\, \oint_{|z_1| =1} \frac{\ud z_1}{2 \pi i z_1} \oint_{|z_2| =1} \frac{\ud z_2}{2 \pi i z_2} \,(1-t_1^{|r_1-r_2|}\,z_1\,z_2^{-1})\,(1-t_2^{|r_2-r_1|}\,z_2\,z_1^{-1}) \label{eq:genHaar}
\eea
While each multiplet's contribution to the index are
\bea
 f_{\rm vector}^{\mathbb{C}^2/\mathbb{Z}_{\infty}}( \vec t,\vec z; \vec r) &=-\frac{2t_1t_2}{1-t_1 t_2}\,\Big(t_1^{|r_1-r_2|}\,z_1\,z_2^{-1}+2+t_2^{|r_2-r_1|}\,z_2\,z_1^{-1}\Big)~, \nn \\
 f_{\rm adjoint}^{\mathbb{C}^2/\mathbb{Z}_{\infty}} ( \vec t,\vec z; \vec r)&=\frac{\sqrt{t_1 t_2}}{1-t_1t_2}\,
 \Big(x^{|r_1-r_2|}\,z_1\,z_2^{-1}+2+x^{|r_2-r_1|}\,z_2\,z_1^{-1}\Big)~, \nn \\ 
 f_{\rm fund/antifund}^{\mathbb{C}^2/\mathbb{Z}_{\infty}} ( \vec t,\vec z; \vec r) &=\frac{\sqrt{t_1 t_2}}{1-t_1t_2} \times \begin{cases} t_2^{|r_1|} z_1+ t_1^{|r_2|}z_2  & \text{fundamental}\\ t_2^{|r_1|} z_1^{-1}+ t_1^{|r_2|}z_2^{-1} & \text{anti-fundamental} \end{cases}~.
\eea

Note that this is in fact the same perturbative contribution as in Section 3.6 of \cite{Gang:2012yr}, which suggest that non-trivial monodromies in the large orbifold limit correspond to insertions of 't Hooft lines in non-trivial representations. Of course, the trivial monodromy case can be thought as no 't Hooft line. Therfore, since in the general case the SUSY preserved by the 't Hooft line is that compatible with the Schur index \cite{Gang:2012yr}, this explains why for no 't Hooft line (or equivalently, $\vec{r}=0$) we recover the Schur index. Note also that supersymmetry then requires the index to depend on a single Lorentz fugacity $\rho=x=\sqrt{t_1\,t_2}$. 

\subsubsection*{Example: $5d$ maximally supersymmetric $U(2)$ gauge theory}
As an explicit test, we can write the large orbifold expression for the perturbative part of the index for the maximally SUSY $U(2)$ theory containing a vector multiplet and an adjoint hyper. Since we have an adjoint hypers, we will have an extra global $SU(2)$ symmetry, under which each chiral in the hyper will have charge, respectively, $1$ and $-1$. Calling the associated fugacity $u$, the perturbative part of the index, together with the appropriate Haar measure, is
\bea
&\int  [\ud \vec z]_{\vec r} \PE \left[ f^{\BC^2/\BZ_{\infty}}_{\vect} (x y, x y^{-1}, \vec z; \vec r) + (u+u^{-1})f^{\BC^2/\BZ_{\infty}}_{\text{adjoint}} (x y, x y^{-1}, \vec z; \vec r)   \right]\nn \\
&= \frac{1}{2}\, \oint_{|z_1| =1} \frac{\ud z_1}{2 \pi i z_1} \oint_{|z_2| =1} \frac{\ud z_2}{2 \pi i z_2} \,(1-x^{|r_1-r_2|}\,z_1\,z_2^{-1})\,(1-x^{|r_2-r_1|}\,z_2\,z_1^{-1})\nn \\
& \qquad {\rm PE}\Big[\,\frac{(u+u^{-1})\,x-2\,x^2}{1-x^2}\,\Big(x^{|r_1-r_2|}\,z_1\,z_2^{-1}+2+x^{|r_2-r_1|}\,z_1^{-1}\,z_2\Big) \,\Big]~,
\eea
where we set $t_1=x y$ and $t_2 = xy^{-1}$.
For $|r_1-r_2|=2$ and $|r_1-r_2|=0$ this is the perturbative part of respectively the first and second lines in Eq. (4.38) of \cite{Gang:2012yr} (see sections \ref{monopolebubbling} and \ref{conclusions} below for the non-perturbative contribution $Z_{\text{mono}}$).

Furthermore, note that both for the adjoint hyper and vector multiplets the corresponding contribution to the indices is proportional to $z^2=z_1\,z_2^{-1}$. In particular, it is clear that the integral will factorize into an integral over $\ud z$ of the $SU(2)$ part times an integral corresponding to a maximally SUSY $U(1)$ theory, \textit{i.e.} a $U(1)$ $\mathcal{N}=4$ vector multiplet in 4d --in fact the integral is trivial since the $U(1)$ adjoint is trivial, so we will simply get an overall factor corresponding to this free multiplet--. Because of this, we can easily find the $SU(2)$ result

\begin{equation}
\oint_{|z| =1} \frac{\ud z}{2 \pi i z}\, (1-x^{|r_1-r_2|}\,z)\,(1-x^{|r_2-r_1|}\,z^{-1})\, {\rm PE}\Big[\,\frac{(u+u^{-1})\,x-2\,x^2}{1-x^2}\,\Big(x^{|r_1-r_2|}\,z+1+x^{|r_2-r_1|}\,z^{-1}\Big) \,\Big]
 \end{equation}
For the particular minimal case when $|r_1-r_2|=1$ this is
\begin{equation}
\oint_{|z| =1} \frac{\ud z}{2 \pi i z}\, (1-x\,z)\,(1-x\,z^{-1})\, {\rm PE}\Big[\,\frac{(u+u^{-1})\,x-2\,x^2}{1-x^2}\,\Big(x\,z+1+x\,z^{-1}\Big) \,\Big]
 \end{equation}
This is  the same result as in Eq. (5.7) of \cite{Gang:2012yr}.

\subsubsection*{General result for $U(N)$ gauge group}
The generalization to the $U(N)$ case with arbitrary action $\vec r = (r_1, \ldots, r_N)$ is now obvious. We assume without loss of generality that
\bea
0 \leq r_1 \leq r_2 \leq \ldots \leq r_{N}~.
\eea
Then, each multiplet's contribution to the index are
\bea
{\rm ind}[D_{SD}({\mathbb{C}^2/\mathbb{Z}_{\infty}})]  &=-\frac{2t_1 t_2}{1-t_1 t_2}\,\Big(N+\sum_{1\leq \alpha<\beta \leq N}  \{ t_1^{|r_\alpha-r_\beta|} z_\alpha\,z_\beta^{-1}+t_2^{|r_\beta-r_\alpha|}\,z_\beta\,z_\alpha^{-1} \} \Big)~, \nn \\
  {\rm ind}[D_{\rm Dirac}({\mathbb{C}^2/\mathbb{Z}_{\infty}})]_{\rm \mathbf{\adj}}&=
 \frac{\sqrt{t_1 t_2}}{1-t_1 t_2} \Big(N+\sum_{1\leq \alpha<\beta \leq N}  \{ t_1^{|r_\alpha-r_\beta|} z_\alpha\,z_\beta^{-1}+t_2^{|r_\beta-r_\alpha|}\,z_\beta\,z_\alpha^{-1} \} \Big)\nn \\ 
{\rm ind}[D_{\rm Dirac}({\mathbb{C}^2/\mathbb{Z}_{\infty}})]_{\Box}  &=\frac{\sqrt{t_1t_2}}{1-t_1 t_2}\,\Big( \sum_{\alpha=1}^N t_2^{r_\alpha}\,z_\alpha\Big)~, \nn \\
{\rm ind}[D_{\rm Dirac}({\mathbb{C}^2/\mathbb{Z}_{\infty}})]_{\overline{\Box}}  &=\frac{\sqrt{t_1t_2}}{1-t_1 t_2}\,\Big(\sum_{\alpha=1}^N t_1^{r_\alpha}\,z_\alpha^{-1}\Big)~.\label{refinedUN}
\eea 
For $SU(N)$ gauge group, we simply impose the condition $\sum_{\alpha=1}^N r_\alpha =0 ~({\rm mod}~n)$, and replace $N$ in the first two equations in \eref{refinedUN} by $N-1$.

In terms of the $x,\, y$ fugacities, and upon appropriately reabsorbing $z_{\alpha}\rightarrow y^{-r_{\alpha}}\,z_{\alpha}$, the contribution of each multiplet is only $x$-dependent, as it should be due to supersymmetry.
\bea
 f_{\rm vector}^{\mathbb{C}^2/\mathbb{Z}_{\infty}}(x ,\vec z; \vec r) &=-\frac{2\,x}{1-x^2}\,\Big(N+\sum_{1\leq \alpha<\beta \leq N}  \{ x^{|r_\alpha-r_\beta|} z_\alpha\,z_\beta^{-1}+x^{|r_\beta-r_\alpha|}\,z_\beta\,z_\alpha^{-1} \} \Big)~, \nn \\
 f_{\rm adjoint}^{\mathbb{C}^2/\mathbb{Z}_{\infty}} (x,\vec z; \vec r)&=
 \frac{x}{1-x^2} \Big(N+\sum_{1\leq \alpha<\beta \leq N}  \{x^{|r_\alpha-r_\beta|} z_\alpha\,z_\beta^{-1}+x^{|r_\beta-r_\alpha|}\,z_\beta\,z_\alpha^{-1} \} \Big)\nn \\ 
 f_{\rm fund}^{\mathbb{C}^2/\mathbb{Z}_{\infty}} (x ,\vec z; \vec r) &=\frac{x}{1-x^2}\,\Big( \sum_{\alpha=1}^N x^{r_\alpha}\,z_\alpha\Big) ~, \nn \\
  f_{\rm antifund}^{\mathbb{C}^2/\mathbb{Z}_{\infty}} (x ,\vec z; \vec r) &=\frac{x}{1-x^2}\,\Big(\sum_{\alpha=1}^N x^{r_\alpha}\,z_\alpha^{-1}\Big) ~.\label{pert}
\eea  

Thus, given a monodromy $\vec r$, the required perturbative part for $U(N)$ gauge theory with matter is
\bea
\CI_p (x, \vec u, \vec z; \vec r) = \PE \left[ f^{\BC^2/\BZ_\infty}_{\vect} (x, \vec z) + \sum_{\vec R} \sum_{i_{\vec R}=1}f^{\BC^2/\BZ_\infty}_{\text{matter};\vec R} (x, \vec z) u_{i_{\vec R}}^{+1/-1}\right]~, \label{eq:Ip}
\eea
where in the second term we sum over all matter present in the theory, $u_{i_{\vec R}}$ denote the flavour fugacities for matter in the representation $\vec R$, and the power $+1/-1$ takes the values $+1$ if $\vec R$ is the fundamental representation and $-1$ for other representations.
Besides, the corresponding generalised  measure is
\begin{equation}
\label{genmeasure}
 \int [ \ud \vec z]_{\vec r} =\frac{1}{N!} \int \prod_{i=1}^N  \frac{\ud z_\alpha}{2\pi i z_\alpha}\,\prod_{1\leq i\leq j\leq N}\, \left(1-x^{|r_\alpha-r_\beta|}\,z_\alpha\,z_\beta^{-1} \right)\left(1-x^{|r_\beta-r_\alpha|}\,z_\beta\,z_\alpha^{-1} \right)~.
\end{equation}
This is in fact the same result as that obtained in \cite{Gang:2012yr}. Thus, all in all, we find that the large orbifold limit of the perturbative part of the 5d index of a gauge theory reduces to the perturbative part of the 4d index with the insertion of a 't Hooft line defect whose charge is given by the monodromy of the 5d gauge bundle $\vec{r}$.

\section{Monopole bubbling indices}\label{monopolebubbling}

In the previous section we have found that the perturbative part of the 5d index on $\mathbb{C}^2/\mathbb{Z}_n$ in the large orbifold limit reproduces the perturbative part of the 4d index with the insertion of a 't Hooft line operator. This suggests that the full 5d index, in the large orbifold limit, can be related to the 't Hooft line index \cite{Ito:2011ea, Gang:2012yr}. A crucial ingredient of the latter is the so-called monopole bubbling effect  \cite{Kapustin:2006pk}, which localizes the 4d 't Hooft line indices on a set of saddle points associated to screening by smooth monopoles. Each such saddle point comes with both a perturbative contribution and a non-perturbative contribution. As discussed above, the large orbifold limit of the 5d index reproduces the perturbative contribution to each saddle point. Unfortunately, the non-perturbative contribution is technically very challenging. Thus, although we cannot obtain cuantitative results, we expect a connection between the large orbifold limit of the 5d non-perturbative contribution to the index and the monopole bubbling index (see section \ref{conclusions} for some speculations about how this might happen).

Inspired nevertheless by this connection with the 4d 't Hooft line index, in this section we leave the 5d realm for a while and focus on the computation of the monopole bubbling contribution to the 't Hooft line index. Recall that for a $U(N)$ gauge theory in the background of the 't Hooft line $T_{\vec B}$ classified by the representation $\vec B$ of $U(N)$, the non-perturbative part of the partition function receives a contribution from certain monopole solutions; this is also known as the monopole bubbling effect, see \eg~ section 10.2 of \cite{Kapustin:2006pk}. For a given $\vec B$, the non-perturbative saddle points are classified by the weights $\vec v$ of the representation $\vec B$. It was pointed out by Kronheimer \cite{kronheimer1985monopoles} that there is a correspondence between such monopole solutions and certain $U(1)$-invariant instanton solutions on a multi-centred Taub-NUT space. The purpose of this section is to explicitly demonstrate Kronheimer's correspondence at the level of partition functions by identifying the monopole bubbling indices, denoted by $Z_{\rm mono}(\vec{B},\vec v)$, with appropriate Hilbert series of instantons on ALE spaces.

Before turning into the detailed computation, let us briefly review the structure of the 't Hooft line index \cite{Ito:2011ea, Gang:2012yr}. For a given representation $\vec B$ of $U(N)$, let us denote by $\mathcal{W}_{\vec B}$ the set of weights of $\vec B$ whose elements are denoted by $\vec v$.  The 't Hooft line index is given by
\begin{equation}
\label{tHooftindex}
\mathcal{I}_{\text{'t Hooft}} (\vec B, \vec v)=\sum_{\vec{v}\in {{\cal W}_{\vec B}}}\,\int [d \vec z]_{\vec{v}}\, \mathcal{I}_{\rm p}(\vec{v})\,Z_{\rm mono}(\vec{B}, \vec v) ~,
\end{equation}
where $[\ud \vec z]_{\vec{v}}$ is the generalised Haar measure given by (\ref{genmeasure}), $\mathcal{I}_{\rm p}(\vec{v})$ is the perturbative part given by \eref{eq:Ip}, and $Z_{\rm mono}(\vec{B},\vec v)$ is the monopole bubbling index.  We have suppressed the dependence on $x$, $\vec z$ and $\vec u$ of the functions $ \mathcal{I}_{\rm p}(\vec{v})$ and $Z_{\rm mono}(\vec{B}, \vec v)$ in the right hand side. We emphasise again that the 't Hooft index depends on both the chosen representation $\vec{B}$ and the chosen weight $\vec{v}$. 

For the highest weight $\vec v= \vec B$, the monopole bubbling index is such that 
\bea Z_{\rm mono}(\vec{B},\,\vec{B})=1~.
\eea
Thus, for the particular case in which we choose $\vec{B}=\vec{0}$, \textit{i.e.} no 't Hooft line the sum in (\ref{tHooftindex}) is absent, so that the non-perturbative contribution associated to monopole bubbling is trivial and, as shown above, both the perturbative contribution and the measure go over to the Schur index and Haar measure respectively, thus recovering the Schur index of \cite{Gadde:2011uv}. In the following we turn to the computation of the monopole bubbling index.

\subsection{Computing monopole bubbling indices}

Let us summarize the computation of monopole bubbling indices presented in \cite{Ito:2011ea, Gang:2012yr}.  The non-perturbative fixed points corresponding to the monopole solutions discussed above are governed by the vector $\vec K = (K_1, \ldots, K_\ell)$ of length $\ell$. According to eq. (5.6) of \cite{Ito:2011ea}, it is related to the Kronheimer's $U(1)$ actions and is determined by the following equation:
\bea
\left(\sum_{\alpha=1}^N g^{B_\alpha} \right)= \left(\sum_{\alpha=1}^N g^{v_\alpha} \right) +( g +g^{-1}-2) \left(\sum_{s=1}^\ell g^{K_s} \right)~, \qquad g \in U(1)~. \label{eqKUN}
\eea

Observe that if $\vec v= \vec B$ (\ie~ $\vec v$ is the highest weight of $\vec B$) or any permutation of $\{B_\alpha: \alpha=1, \ldots, N\}$, then this equation admits no solution for $\vec K$ and the contribution from the monopole bubbling is trivial, $Z_{\rm mono}(\vec{B},\,\vec{v})=1$.  A representation $\vec{B}$ of which all of its weights $\vec v$ are permutations of $\{B_\alpha\}$ is referred to as a {\it minuscule representation}; thus, for such a $\vec B$, $Z_{\rm mono}(\vec{B},\,\vec{v})=1$ for all $\vec v \in {\cal W}_{\vec B}$.

In order to compute the monopole bubbling indices, we consider the $N$-tuple of Young diagrams $\vec Y = (Y_1, \ldots, Y_N)$ that satisfy all of the following conditions:
\ben
\item The total number of boxes must equal to the length $\ell$ of vector $\vec K$:
\bea
\sum_{\alpha=1}^N |Y_\alpha| = \ell~.
\eea
\item Upon assigning the numbers $v_{\alpha(s)}+j_{\alpha(s)}-i_{\alpha(s)}$ to each box $\alpha(s)$ located at the $i_{\alpha(s)}$-th row and $j_{\alpha(s)}$-th column in the Young diagram $Y_\alpha$, we select only $\vec Y$ such that the following equality is satisfied:
\bea
K_s = v_{\alpha(s)}+j_{\alpha(s)}-i_{\alpha(s)}~, \quad \text{for all $s \in Y_{\alpha}$}~.
\eea
\een
We denote by $\CR(\vec B, \vec v; \vec K)$ the set of all $N$-tuples of Young diagrams satisfying the above conditions.

Once the relevant Young diagrams have been identified, the contributions from the vector multiplet and hypermultiplets can be derived from \eref{eq:HSC2vec}, \eref{eq:HSC2fund}, \eref{eq:HSC2antifund} and \eref{eq:HSC2adjoint}, with the monomials in the $\PE$s projected such that only terms that are invariant under the following transformations are kept:
\bea
t_1 \rightarrow g^{-1} t_1, \qquad t_2 \rightarrow g t_2, \qquad z_\alpha \rightarrow g^{v_\alpha} z_\alpha~, \qquad g \in U(1)~.
\eea 
Explicitly,
{\small
\bea
&Z_{\text{vector}} (t_1, t_2, \vec z; \vec Y; \vec v) \nn \\
&  =   \PE \Bigg[ \sum_{\alpha, \beta=1}^N \sum_{s \in Y_\alpha } \left(\frac{z_\alpha}{z_\beta} t_1^{-l_{Y_\beta}(s)}t_2^{1+a_{Y_\alpha }(s)}+ \frac{z_\beta}{z_\alpha} t_1^{1+l_{Y_\beta }(s)}t_2^{-a_{Y_\alpha}(s)} \right) \delta_{a_{Y_\alpha}(s)-l_{Y_\beta}(s)-1, v_\alpha-v_\beta} \Bigg] ~,  \\ 
& Z_{\text{fund}} (t_1, t_2, \vec z,  u; \vec Y; \vec v) =   \PE \Bigg[ u^{-1}  \sum_{\alpha=1}^N z_\alpha\sum_{(a,b) \in Y_\alpha }   t_1^a t_2^b~ \delta_{v_\alpha-a+b,0}\Bigg] ~,  \\
& Z_{\text{antifund}} (t_1, t_2, \vec z, u; \vec Y; \vec v) =\PE \left[ u  \sum_{\alpha=1}^N z_\alpha  \sum_{(a,b) \in Y_\alpha }  t_1^{a-1} t_2^{b-1} \delta_{v_\alpha-a+b,0}  \right] ~, \\
& Z_{\text{adjoint}} (t_1, t_2, \vec z,u ; \vec Y; \vec v) \nn \\
& = \PE \left[ u\sum_{\alpha, \beta=1}^N \sum_{s \in Y_\alpha } \left(\frac{z_\alpha}{z_\beta} t_1^{-l_{Y_\beta}(s)}t_2^{1+a_{Y_\alpha }(s)}  + \frac{z_\beta}{z_\alpha} t_1^{1+l_{Y_\beta }(s)}t_2^{-a_{Y_\alpha}(s)} \right) \delta_{a_{Y_\alpha}(s)-l_{Y_\beta}(s)-1, v_\alpha-v_\beta} \right]~.
\eea}
where $u$ denotes the fugacity for the flavour symmetry.  Note that in order to find an agreement with the results  in \cite{Ito:2011ea, Gang:2012yr}, we need to make the following redefinitions for the flavour fugacities:
\bea
\begin{array}{lll}
u &\rightarrow u\sqrt{t_1t_2} &\qquad \text{for the fundamental hypermultiplet} \\ 
u &\rightarrow u^{-1} \sqrt{t_1t_2} &\qquad \text{for the anti-fundamental hypermultiplet} \\ 
u &\rightarrow \frac{u}{\sqrt{t_1t_2}} &\qquad \text{for the adjoint hypermultiplet}~.
\end{array}
\eea
In this section, we adopt such redefinitions.

Finally, the monopole bubbling index for the $U(N)$ gauge theory is given by
\bea
Z^{U(N)}_{\text{mono}}(\vec B, \vec v)  (\vec t, \vec z, \vec u)  = \sum_{\vec Y \in \CR( \vec B, \vec v)} \frac{Z_{\text{vector}} (\vec t, \vec z; \vec Y) }{Z_{\text{matter}} (\vec t, \vec z, \vec u; \vec Y)}~.
\eea

\subsection{Monopole bubbling indices for pure $U(N)$ theories and the Hilbert series of instantons in gauge theory on $\BC^2/\BZ_n$} \label{sec:monoVShs}

We now concentrate on the case of pure gauge theories. In this case, we make the following observation on the relation between the monopole bubbling index for $U(N)$ gauge theory and the Hilbert series of instantons in $SU(N)$ gauge theory on $A$-type ALE space \cite{DHMRS2013}.
 \begin{quote}
Given a representation $\vec{B}=(n_1,n_2,\ldots,n_N)$ of $U(N)$ and a weight $\vec v$ of $\vec B$, the corresponding monopole bubbling index $Z^{U(N)}_{\text{mono}} (\vec B, \vec v)$ {\it is equal to} the Hilbert series of instantons in $SU(N)$ gauge theory on $\BC^2/\BZ_n$, where $n =\sum_{\alpha=1}^N |n_\alpha|$,  with the holonomy $\vec {r} =\vec{v}$ and with the Kronheimer-Nakajima vector $\vec k$ such that $k_j$ (with $j = 1,...,n$) is the number of times that the number $j~(\mathrm{mod}~n)$ appear in the vector $\vec K$ in \eref{eqKUN}.
\end{quote}

For the representation $\vec{B}=(n,0,\ldots,0)$ of $U(N)$, the corresponding Hilbert series is that of {\bf pure}\footnote{Recall that by `pure instanton', we mean the instanton bundle with vanishing first Chern class: $\beta_1= \beta_2 = \cdots = \beta_{n-1}=0$. This is not to be confused with instantons in a pure gauge theory.} $SU(N)$ instanton on $\BC^2/\BZ_n$ with the monodromy $\vec{v}$.

Note that in the special case of $U(2)$, the representation $\vec{B}=(n,0)$ of $U(2)$ can also be identified with the $n+1$ dimensional (or spin $n/2$) representation of $SU(2)$.  For a given $\vec v=(p,n-p)$, with $p=0, \ldots, n$, the corresponding pure $SU(2)$ instantons on $\BC^2/\BZ_n$ has an instanton number $k = p(n-p)/n$.

It is important to stress that the ALE space on which the instanton whose Hilbert series captures the monopole bubbling index lives should be viewed merely as {\it an auxiliary device}, as in \cite{kronheimer1985monopoles} (see also the Appendix C of  \cite{Ito:2011ea}). Hence such an ALE space should not to be confused with the physical orbifold target space in which the 5d theory considered in section \ref{orbifoldindex} and \ref{largeorbifold} lives. Indeed, in the case at hand the orbifold degree corresponds to the 't Hooft monopole charge, so that the large orbifold limit corresponds simply to a large charge monopole; this is in contrast to the $5d\; \rightarrow \; 4d$ reduction in section \ref{largeorbifold}. See appendix \ref{largeB} for monopole bubbling indices of large charge 't Hooft operators.

Let us now turn to some specific examples.

\subsubsection{$\CN=2$ pure $U(2)$ gauge theory: $\vec B=(2,0)$} \label{sec:N2pureU2B20}

Given the representation $\vec B=(2,0)$ of $U(2)$, the weights $\vec v$ are $(2,0)$, $(1,1)$ and $(0,2)$.  

For $\vec v=(2,0)$ or $\vec v =(0,2)$, we obtain from \eref{eqKUN}
\bea
\sum_{s=1}^\ell x^{K_s} = 0~;
\eea
and so there is no solution $\vec K$ for these $\vec v$.  Thus,
\bea
Z_{\text{mono}} (\vec B=(2,0), \vec v=(2,0)) = Z_{\text{mono}} (\vec B=(2,0), \vec v=(0,2)) =1~.
\eea  
This agrees with the Hilbert series of $SU(2)$ instanton on $\BC^2/\BZ_2$ with $\vec r =(2,0) \equiv (0,2) \equiv (0,0)$ modulo $2$, and $\vec k =(0,0)$.

For $\vec v =(1,1)$, the solution to \eref{eqKUN} is
\bea
\vec K = (1)~.
\eea
The corresponding set $\CR(\vec B=(2,0), \vec v =(1,1); \vec K=(1)) $ is given by
\bea
\CR = \{ (\emptyset, \Box), (\Box, \emptyset) \}~. \label{setRK1v11}
\eea
The monopole bubbling index receives only the vector multiplet contribution:
\bea
&Z_{\text{mono}} (\vec B=(2,0), \vec v=(1,1)) (\vec t, \vec z)  \nn \\
&= \sum_{\vec Y \in \CR} Z_{\text{vector}} (\vec t, \vec z; \vec Y; \vec v = (1,1)) \nn \\
&= \frac{1}{\left(1-\frac{t_1 t_2 z_1}{z_2}\right) \left(1-\frac{z_2}{z_1}\right)}+\frac{1}{\left(1-\frac{z_1}{z_2}\right) \left(1-\frac{t_1 t_2 z_2}{z_1}\right)} \nn \\
&= \frac{1-(t_1 t_2)^2}{(1- t_1 t_2)(1-t_1t_2 z_1 z_2^{-1})(1-t_1t_2 z_1^{-1} z_2) } ~. \label{Zmono2011pureU2}
\eea
This is the Hilbert series of $\BC^2/\BZ_2$.  This agrees with the Hilbert series of $1/2$ pure $SU(2)$ instanton on $\BC^2/\BZ_2$ with $\vec r =(1,1)$ and $\vec k =(1,0)$.

\subsubsection{$\CN=2$ pure $U(2)$ gauge theory:  $\vec B=(3,0)$} \label{sec:N2pureU2B30}

Given the representation $\vec B=(3,0)$ of $U(2)$, the weights $\vec v$ are 
\bea (3,0), \quad (1,2), \quad  (2,1), \quad (0,3)~. \eea 
For $\vec v=(3,0)$ or $\vec v=(0,3)$, there is no solution $\vec K$ in \eref{eqKUN}.  Hence,
\bea
Z_{\text{mono}} (\vec B=(3,0), \vec v=(3,0)) = Z_{\text{mono}} (\vec B=(3,0), \vec v=(0,3)) =1~.
\eea 
This agrees with the Hilbert series of $SU(2)$ instanton on $\BC^2/\BZ_3$ with $\vec r =(3,0) \equiv (0,3) \equiv (0,0)$ modulo $3$, and $\vec k =(0,0)$.

For $\vec v =(1,2)$ or $\vec v=(2,1)$, there are two solutions: $\vec K=(1,2)$ and $\vec K =(2,1)$.  For each of such $\vec v$, one has to sum over both solutions $\vec K$.

For $\vec v=(1,2)$, the sets $\CR( \vec B, \vec v; \vec K)$ are given by
\bea
\CR(\vec B=(3,0), \vec v =(1,2); \vec K =(1,2)) &= \{ ((1),(1)), ((2), \emptyset) \} \label{setRB30v12K12} \\
\CR(\vec B=(3,0), \vec v =(1,2); \vec K =(2,1)) &= \{ ( \emptyset, (1,1) ) \}  \label{setRB30v12K21} ~.
\eea
 The monopole bubbling index is then given by
\bea
&Z_{\text{mono}} (\vec B=(3,0), \vec v=(1,2)) (\vec t, \vec z)  \nn \\
&= \sum_{\vec K =(1,2), (2,1)} \sum_{\vec Y \in \CR((3,0), (1,2); \vec K)} Z_{\text{vector}} (\vec t, \vec z; \vec Y; \vec v = (1,2)) \nn \\
&= \frac{1}{\left(1-\frac{t_2 z_1}{z_2}\right) \left(1-\frac{t_1 z_2}{z_1}\right)}+\frac{1}{\left(1-\frac{t_1 t_2^2 z_1}{z_2}\right) \left(1-\frac{z_2}{t_2 z_1}\right)}+\frac{1}{\left(1-\frac{z_1}{t_1 z_2}\right) \left(1-\frac{t_1^2 t_2 z_2}{z_1}\right)}~.
\eea
Setting $t_1 = x y, t_2= x y^{-1}$, we find that
\bea
&Z_{\text{mono}} (\vec B=(3,0), \vec v=(1,2)) (x, x, \vec z) \nn \\
&= 1+t^2+\left(\frac{z_1}{y z_2}+\frac{y z_2}{z_1}\right) t^3+t^4+\left(\frac{z_1}{y z_2}+\frac{y z_2}{z_1}\right) t^5+\left(1+\frac{z_1^2}{y^2 z_2^2}+\frac{y^2 z_2^2}{z_1^2}\right) t^6 \nn \\
& \quad +\left(\frac{z_1}{y z_2}+\frac{y z_2}{z_1}\right) t^7+\left(1+\frac{z_1^2}{y^2 z_2^2}+\frac{y^2 z_2^2}{z_1^2}\right) t^8+\ldots  \nn \\
&= g_{\BC^2/\BZ_3} (x, y^{-1/3} z_1^{1/3} z_2^{-1/3}) \label{HSC2Z3}
\eea
where the Hilbert series of $\BC^2/\BZ_3$ is given by
\bea
g_{\BC^2/\BZ_3} (t,z) = \frac{1}{3} \sum_{j=0}^2 \frac{1}{ (1-\omega^j t z)(1- \omega^{-j} t z^{-1}) }~, \qquad \omega^3=1~.
\eea
This agrees with the Hilbert series of $2/3$ pure $SU(2)$ instanton on $\BC^2/\BZ_3$ with $\vec r =(1,2)$, and $\vec k =(1,1,0)$.  Setting $z_1=z_2=1$, we obtain
\bea
&Z_{\text{mono}} (\vec B=(3,0), \vec v=(1,2)) (x, x, 1,1)  \nn \\
&= 1+x^2+2 x^3+x^4+2 x^5+3 x^6+2 x^7 +3 x^8+4 x^9+3 x^{10}+\ldots~.
\eea

Similarly, for $\vec v=(2,1)$,  it can be shown that
\bea
Z_{\text{mono}} (\vec B=(3,0), \vec v=(2,1)) (t_1,t_2, \vec z) = Z_{\text{mono}} (\vec B=(3,0), \vec v=(1,2)) (t_2,t_1, \vec z)~. 
\eea

\subsubsection{$\CN=2$ pure $U(2)$ gauge theory: $\vec B=(4,0)$} \label{sec:N2pureU2B40}

The weights of $\vec B=(4,0)$ are $(4,0)$, $(3,1)$, $(2,2)$, $(1,3)$ and $(0,4)$.
Similarly to the previous examples, we know that
\bea
& Z_{\text{mono}} (\vec B=(4,0), \vec v=(4,0)) = Z_{\text{mono}} (\vec B=(4,0), \vec v=(0,4)) =1~, \nn \\
& Z_{\text{mono}} (\vec B=(4,0), \vec v=(3,1)) (t_1,t_2,\vec z) = Z_{\text{mono}} (\vec B=(4,0), \vec v=(1,3)) (t_2,t_1,\vec z)~.
\eea

For $\vec v =(3,1)$, the corresponding $\vec K$ are $(1,2,3)$ and its permutations.  The sets $ \CR((4,0), (3,1); \vec K)$ are
\bea
&\vec K =(1,2,3): \qquad \{ (\emptyset, (3)) \}~, \nn \\
&\vec K = (3,1,2):\qquad \{((1),(2))\}~, \nn \\
&\vec K=(3,2,1): \qquad \{((1,1),(1)),((1,1,1),\emptyset)\}~;
\eea
other $\vec K$ give rise to empty sets.
The monopole bubbling index for $\vec v =(3,1)$ is then given by
\bea
&Z_{\text{mono}} (\vec B=(4,0), \vec v=(3,1)) (\vec t, \vec z)  \nn \\
&= \sum_{\vec K: \; \text{perms$(1,2,3)$}} \quad \sum_{\vec Y \in \CR((4,0), (3,1); \vec K)} Z_{\text{vector}} (\vec t, \vec z; \vec Y; \vec v = (3,1)) \nn \\
&= \frac{1}{\left(1-\frac{t_1^3 t_2 z_1}{z_2}\right) \left(1-\frac{z_2}{t_1^2 z_1}\right)}+\frac{1}{\left(1-\frac{t_1^2 z_1}{z_2}\right) \left(1-\frac{t_2 z_2}{t_1 z_1}\right)}+\frac{1}{\left(1-\frac{t_1 z_1}{t_2 z_2}\right) \left(1-\frac{t_2^2 z_2}{z_1}\right)} \nn \\
& \quad +\frac{1}{\left(1-\frac{z_1}{t_2^2 z_2}\right) \left(1-\frac{t_1 t_2^3 z_2}{z_1}\right)}~.
\eea
Setting $t_1 = x y, t_2= x y^{-1}$, we find that
\bea
&Z_{\text{mono}} (\vec B=(4,0), \vec v=(3,1)) (xy, xy^{-1}, \vec z)   \nn \\
&= 1+x^2+\left(1+\frac{y^2 z_1}{z_2}+\frac{z_2}{y^2 z_1}\right) x^4+\left(1+\frac{y^2 z_1}{z_2}+\frac{z_2}{y^2 z_1}\right) x^6 \nn \\
& \quad +\left(1+\frac{y^4 z_1^2}{z_2^2}+\frac{y^2 z_1}{z_2}+\frac{z_2}{y^2 z_1}+\frac{z_2^2}{y^4 z_1^2}\right) x^8+\left(1+\frac{y^4 z_1^2}{z_2^2}+\frac{y^2 z_1}{z_2}+\frac{z_2}{y^2 z_1}+\frac{z_2^2}{y^4 z_1^2}\right) x^{10} + \ldots \nn \\
&= g_{\BC^2/\BZ_4} (x, y^{1/2} z_1^{1/4} z_2^{-1/4}) \label{B40v31}
\eea
where the Hilbert series of $\BC^2/\BZ_4$ is given by
\bea
g_{\BC^2/\BZ_4} (t,z) = \frac{1}{4} \sum_{j=0}^3 \frac{1}{ (1-\omega^j t z)(1- \omega^{-j} t z^{-1}) }~, \qquad \omega^4=1~.
\eea
This agrees with the Hilbert series of $3/4$ pure $SU(2)$ instanton on $\BC^2/\BZ_4$ with $\vec r =(3,1)$, and $\vec k =(1,1,1,0)$.

For $\vec v =(2,2)$, the corresponding $\vec K$ are $(1,2,2,3)$ and its permutations. The computation is similar to the previous example.  We find that the monopole bubbling index can be written in terms of an $SU(2)$ character expansion:
\bea \label{B40v22}
&Z_{\text{mono}} (\vec B=(4,0), \vec v=(2,2)) (xy, xy^{-1}, z,1/z)  \nn \\
&= \frac{1}{1-t^4} \left( [2m_2 +2m_4]_{z} x^{2m_2+4m_4} + [2m_2 +2m_4+2]_{z} x^{2m_2+4m_4+6}  \right)~,
\eea
where $[a]_z$ denotes the character of the $SU(2)$ representation $[a]$ in terms of the variable $z$.
Observe that this does not depend on the fugacity $y$. This is in fact the Hilbert series of $1$ $SU(2)$ instantons on $\BC^2/\BZ_4$ with $\vec r = (2,2)$ and $\vec k =(1,2,1,0)$.  The unrefined index is 
\bea
Z_{\text{mono}} (\vec B=(4,0), \vec v=(2,2)) (x, x, 1,1)   = \frac{1+x^2+2 x^4+x^6+x^8}{\left(1-x^2\right)^4 \left(1+x^2\right)^2}~.
\eea

\subsubsection{$\CN=2$ pure $U(3)$ gauge theory} 
The computations for the pure $U(3)$ gauge theory are similar to the preceding section.  Let us summarise the matchings between the monopole bubbling indices and the Hilbert series of instantons on $\BC^2/\BZ_n$ in \tref{tab:matchU3SYM}.
\begin{table}[H]
\begin{center}
\begin{tabular}{|c|c||c|c|c|c|}
\hline
\multicolumn{2}{|c||}{Monopole bubbling} & \multicolumn{4}{|c|}{Hilbert series of instantons} \\
\hline
$\vec B$ & $\vec v$ & Description & $\vec r$  & $\vec k$ & Hilbert series \\
\hline
$(2,0,0)$ & $(1,1,0)$ & $1/2$ $SU(3)$ pure inst., $\BC^2/\BZ_2$ & $(1,1,0)$ & $(1,0)$ & $\BC^2/\BZ_2$, \eref{Zmono2011pureU2} \\
\hline
$(3,0,0)$ &  $(2,1,0)$ & $2/3$ $SU(3)$ pure inst., $\BC^2/\BZ_3$ & $(2,1,0)$ & $(1,1,0)$ & $\BC^2/\BZ_3$, \eref{HSC2Z3} \\
$(3,0,0)$ &  $(1,1,1)$ & $1$ $SU(3)$ pure inst., $\BC^2/\BZ_3$ & $(1,1,0)$ & $(2,1,0)$ & \large{$\frac{1+2 x^2+2 x^4+x^6}{\left(1-x^2\right)^6}$} \\
\hline
$(2,1,0)$ & $(1,1,1)$ & $SU(3)$ non-pure inst., $\BC^2/\BZ_3$ & $(1,1,1)$ & $(1,0,0)$ & $\widetilde{\CM}_{\text{1,$SU(3)$,$\BC^2$}}$, (2.1) of \cite{Benvenuti:2010pq}  \\
 & & & & & \large{$\frac{1+4 x^2+x^4}{\left(1-x^2\right)^4}$} \\
\hline
$(2,2,0)$ & $(2,1,1)$ & $SU(3)$ non-pure inst., $\BC^2/\BZ_4$ & $(2,1,1)$ & $(1,0,0,0)$ & $\BC^2/\BZ_2$, \eref{Zmono2011pureU2}\\
\hline
\end{tabular}
\end{center}
\caption{Matchings between the monopole bubbling indices for $4d$ $\CN=2$ $U(3)$ pure gauge theory in the background of the 't Hooft line $T_B$ and the Hilbert series of $SU(3)$ instantons on $\BC^2/\BZ_n$. In the above, $\widetilde{\CM}_{\text{1,$SU(3)$,$\BC^2$}}$ denotes the reduced instanton moduli space of one $SU(3)$ instanton on $\BC^2$.}
\label{tab:matchU3SYM}
\end{table}

\subsection{Adding matter}

By restricting to the simplest case of pure gauge theories we have found a nice characterization of the monopole bubbling indices as Hilbert series of certain instantons. We now extend this characterization to theories with matter.

\subsubsection{$\CN=2^*$ $U(2)$ gauge theory: $\vec B=(2,0)$} \label{sec:N2starU220}

Similarly to Section \ref{sec:N2pureU2B20}, we find that for $\vec v =(2,0)$ and $\vec v =(0,2)$ the monopole bubbling index is given by
\bea
Z_{\text{mono}} (\vec B=(2,0), \vec v=(2,0)) = Z_{\text{mono}} (\vec B=(2,0), \vec v=(0,2)) =1~.
\eea  
For $v=(1,1)$, the set $\CR$ of ordered pairs of the Young diagrams is given by \eref{setRK1v11}. The corresponding monopole bubbling index is
\bea
& Z_{\text{mono}} (\vec B=(2,0), \vec v=(1,1)) (u, \vec t, \vec z)  \nn \\
&= \sum_{\vec Y \in \CR} \frac{Z_{\text{vector}} (\vec t, \vec z; \vec Y; \vec v = (1,1))}{Z_{\text{adjoint}} (\vec t, \vec z, u; \vec Y; \vec v = (1,1))} \nn \\
 &= \frac{\left(1-\frac{u \sqrt{t_1} \sqrt{t_2} z_1}{z_2}\right) \left(1-\frac{u z_2}{\sqrt{t_1} \sqrt{t_2} z_1}\right)}{\left(1-\frac{t_1 t_2 z_1}{z_2}\right) \left(1-\frac{z_2}{z_1}\right)}+\frac{\left(1-\frac{u z_1}{\sqrt{t_1} \sqrt{t_2} z_2}\right) \left(1-\frac{u \sqrt{t_1} \sqrt{t_2} z_2}{z_1}\right)}{\left(1-\frac{z_1}{z_2}\right) \left(1-\frac{t_1 t_2 z_2}{z_1}\right)}~.
\eea
Indeed, this is in agreement with the instanton computation \eref{5dN4SYM1011} on $\BC^2/\BZ_2$, with $\vec k =(1,0)$ and $\vec r=(1,1)$, after redefining $u \rightarrow \frac{u}{\sqrt{t_1t_2}}$ in the latter.

\paragraph{Comparison with \cite{Gang:2012yr}.} 
The $5d$ partition function \eref{5dN4SYM1011} can be equated with (4.44) of \cite{Gang:2012yr} by redefining $u \rightarrow \frac{u}{\sqrt{t_1 t_2}}$ and multiplying by an overall factor:
\bea
 x u^{-1} H^{\BC^2/\BZ_2}_{\text{inst}; \vec k=(1,0), \vec r=(1,1)} (x,x,\vec z, u x^{-1})  = \frac{1+[1]_u x -2 [2]_{\vec z} x^2+[1]_u x^3+x^4}{(1- x^2 z_1 z_2^{-1})(1- x^2 z_2 z_1^{-1})}~,
\eea
where $t_1 t_2 =x^2$.
In a similar way, \eref{4dN4SYM1011} can be equated with (4.36) of \cite{Gang:2012yr} by shifing the mass parameter by $-\epsilon_+/2$:\footnote{The $\Omega$-deformation parameters $\epsilon_{1,2}$ are set to $\rho$ in \cite{Gang:2012yr}.}
\bea
& Z^{\BC^2/\BZ_2}_{\text{inst};  \vec k=(1,0), \vec r=(1,1)} (\epsilon_1, \epsilon_2, \vec a,  \mu-\epsilon_+/2) \nn \\
&= \frac{\left(a_1-a_2+\mu - \epsilon_+/2\right) \left(-a_1+a_2+\mu+\epsilon _+/2\right)}{\left(a_1-a_2\right) \left(-a_1+a_2+\epsilon_+\right)} + (\epsilon_i \rightarrow -\epsilon_i)~.
\eea

\subsubsection{$\CN=2^*$ $U(2)$ gauge theory: $\vec B=(3,0)$} \label{sec:N2starU230}

We proceed in a similar way to Section \ref{sec:N2pureU2B30}. The monopole bubbling indices for $\vec v =(3,0)$ and $\vec v =(0,3)$ are given by
\bea
&Z_{\text{mono}} (\vec B=(3,0), \vec v=(3,0)) = Z_{\text{mono}} (\vec B=(3,0), \vec v=(0,3)) =1~.
\eea
For $\vec v =(1,2)$, the relevant sets $\CR((3,0), (1,2); \vec K)$, with $\vec K$ being $(1,2)$ or $(2,1)$, are given by \eref{setRB30v12K12} and \eref{setRB30v12K21}.  The corresponding monopole bubbling index is
\bea
&Z_{\text{mono}} (\vec B=(3,0), \vec v=(1,2)) (u, \vec t, \vec z)  \nn \\
&= \sum_{\vec K =(1,2), (2,1)} \quad \sum_{\vec Y \in \CR((3,0), (1,2); \vec K)} \frac{Z_{\text{vector}} (\vec t, \vec z; \vec Y; \vec v = (1,2))}{Z_{\text{adjoint}} (\vec t, \vec z, u; \vec Y; \vec v = (1,2))} \nn \\
&=\frac{\left(1-\frac{u t_1^{3/2} \sqrt{t_2} z_1}{z_2}\right) \left(1-\frac{u z_2}{t_1^{3/2} \sqrt{t_2} z_1}\right)}{\left(1-\frac{t_1^2 t_2 z_1}{z_2}\right) \left(1-\frac{z_2}{t_1 z_1}\right)}+\frac{\left(1-\frac{u \sqrt{t_1} z_1}{\sqrt{t_2} z_2}\right) \left(1-\frac{u \sqrt{t_2} z_2}{\sqrt{t_1} z_1}\right)}{\left(1-\frac{t_1 z_1}{z_2}\right) \left(1-\frac{t_2 z_2}{z_1}\right)} \nn \\
& \qquad +\frac{\left(1-\frac{u z_1}{\sqrt{t_1} t_2^{3/2} z_2}\right) \left(1-\frac{u \sqrt{t_1} t_2^{3/2} z_2}{z_1}\right)}{\left(1-\frac{z_1}{t_2 z_2}\right) \left(1-\frac{t_1 t_2^2 z_2}{z_1}\right)}
\eea
Indeed, this is in agreement with the instanton computation \eref{5dN2starr12} on $\BC^2/\BZ_3$, with $\vec k =(1,1,0)$ and $\vec r =(1,2)$, upon a rescaling $u \rightarrow \frac{u}{\sqrt{t_1 t_2}}$ in the latter.

Similarly, it can be shown that
\bea
Z_{\text{mono}} (\vec B=(3,0), \vec v=(2,1)) (u, t_1,t_2, \vec z) = Z_{\text{mono}} (\vec B=(3,0), \vec v=(1,2)) (u, t_2,t_1, \vec z)~. 
\eea
 
\paragraph{Comparison with \cite{Gang:2012yr}.} Setting $t_1=t_2=x$, we find that
\bea
&u^{-1}x Z_{\text{mono}} (\vec B=(3,0), \vec v=(1,2)) (u, x, x, \vec z) \nn \\
&= \frac{\left(\frac{1}{u}+u\right) \left(x+x^3+x^5\right)+2 \left(1-x^2-x^4+x^6\right)-3 x^3 \left(\frac{z_1}{z_2}+\frac{z_2}{z_1}\right)}{\left(1-\frac{x^3 z_1}{z_2}\right) \left(1-\frac{x^3 z_2}{z_1}\right)} \nn \\
&=2+\left(\frac{1}{u}+u\right) x-2 x^2+\left(\frac{1}{u}+u-\frac{z_1}{z_2}-\frac{z_2}{z_1}\right) x^3+\left(-2+\frac{z_1}{u z_2}+\frac{u z_1}{z_2}+\frac{z_2}{u z_1}+\frac{u z_2}{z_1}\right) x^4 \nn \\
& \qquad +\left(\frac{1}{u}+u-\frac{2 z_1}{z_2}-\frac{2 z_2}{z_1}\right) x^5+ \ldots~.
\eea
This is in agreement with (4.45) of \cite{Gang:2012yr}.

\subsubsection{$\CN=2$ $U(2)$ gauge theory with $4$ flavours:  $\vec B=(1,-1)$} \label{sec:U24flvB20}
The vector $\vec B =(1,-1)$ corresponds to the adjoint representation with the weights $\vec v$
\bea
(1, -1), \qquad (0,0), \qquad (-1,1)~.
\eea
For $\vec v=(1,-1)$ and $\vec v=(-1,1)$, 
\bea
Z_{\text{mono}} (\vec B=(1,-1), \vec v=(1,-1)) = Z_{\text{mono}} (\vec B=(1,-1), \vec v=(-1,1)) =1~.
\eea  

For $\vec v=(0,0)$, the solution to \eref{eqKUN} is $\vec K=(0)$ and hence the corresponding set $\CR(\vec B=(1,-1), \vec v=(0,0); \vec K=(0))$ of ordered pairs of the Young diagrams is given by \eref{setRK1v11}. The corresponding monopole bubbling index is
\bea
& Z_{\text{mono}} (\vec B=(1,-1), \vec v=(0,0)) (\vec u, \vec t, \vec z)  \nn \\
&= \sum_{\vec Y \in \CR} \frac{Z_{\text{vector}} (\vec t, \vec z; \vec Y; \vec v = (0,0))}{\prod_{i=1}^2 Z_{\text{antifund}} (\vec t, \vec z, u_i; \vec Y; \vec v = (0,0))\prod_{j=3}^4 Z_{\text{fund}} (\vec t, \vec z, u_j; \vec Y; \vec v = (0,0))} \nn \\
&= \frac{\left(1-\frac{\sqrt{t_1} \sqrt{t_2} z_1}{u_1}\right) \left(1-\frac{\sqrt{t_1} \sqrt{t_2} z_1}{u_2}\right) \left(1-\frac{\sqrt{t_1} \sqrt{t_2} z_1}{u_3}\right) \left(1-\frac{\sqrt{t_1} \sqrt{t_2} z_1}{u_4}\right)}{\left(1-\frac{t_1 t_2 z_1}{z_2}\right) \left(1-\frac{z_2}{z_1}\right)} \nn\\
& \qquad +\frac{\left(1-\frac{\sqrt{t_1} \sqrt{t_2} z_2}{u_1}\right) \left(1-\frac{\sqrt{t_1} \sqrt{t_2} z_2}{u_2}\right) \left(1-\frac{\sqrt{t_1} \sqrt{t_2} z_2}{u_3}\right) \left(1-\frac{\sqrt{t_1} \sqrt{t_2} z_2}{u_4}\right)}{\left(1-\frac{z_1}{z_2}\right) \left(1-\frac{t_1 t_2 z_2}{z_1}\right)}~.
\eea
Indeed, this is in agreement with the instanton computation \eref{U24flvk10r11} on $\BC^2/\BZ_2$, with $\vec k =(0,1)$ and $\vec r =(0,0)$, upon the following rescalings in the latter:
\bea
\begin{array}{lll}
u_i &\rightarrow  u_i^{-1}\sqrt{t_1 t_2}~, &\qquad i=1,2~, \\
u_j &\rightarrow u_j \sqrt{t_1 t_2}~, &\qquad j=3,4~.
\end{array}
\eea

Setting $t_i = e^{-\beta \epsilon_i}$, $u_j = e^{-\beta \mu_j}$ and $z_\alpha = e^{-\beta a_\alpha}$ and taking limit $\beta \rightarrow 0$, we have  
\bea
&\lim_{\beta \rightarrow 0} \beta^{-2} Z_{\text{mono}} (e^{-\beta \vec \mu}, e^{-\beta \vec \epsilon},  e^{-\beta \vec a}) \nn \\
&=\frac{\left(2 a_1+\epsilon _1+\epsilon _2-2 \mu _1\right) \left(2 a_1+\epsilon _1+\epsilon _2-2 \mu _2\right) \left(2 a_1+\epsilon _1+\epsilon _2-2 \mu _3\right) \left(2 a_1+\epsilon _1+\epsilon _2-2 \mu _4\right)}{16 \left(-a_1+a_2\right) \left(a_1-a_2+\epsilon _1+\epsilon _2\right)} \nn \\
& \quad + (a_1 \leftrightarrow a_2)~;
\eea 
this is in agreement with the last line of Eq. (6.17) in \cite{Ito:2011ea}.

\section{Conclusions and speculations}\label{conclusions}

In this paper we have studied indices for 5d theories on orbifold backgrounds of the form $S^1\times S^4/\mathbb{Z}_n$. The 5d index contains both a perturbative and a non-perturbative contribution, whose orbifold version we have considered in Section \ref{orbifoldindex}. 

Since the space where the theories under consideration are placed contains two circles, namely the ``time" $S^1$ and the orbifolded circle of $S^4/\mathbb{Z}_n$, we can imagine dimensionally reducing along either of them. Dimensionally reducing along the ``time" $S^1$ leads to the partition function on $S^4/\mathbb{Z}_n$ of the 4d version of the theory. Such dimensional reduction is implemented by the standard Nekrasov limit $\beta\rightarrow 0$ on the 5d index. In turn, we can implement the dimensional reduction along the orbifolded direction by taking the large orbifold limit. Note that, since this procedure does not involve the ``time" circle where the supersymmetric boundary conditions are imposed, the resulting quantity must be an index. Indeed, we find evidence that such dimensional reduction leads to the index of the 4d reduction of the theory in the presence of a 't Hooft line. While such result is robust for the perturbative sector (see section \ref{sec:pertinforb}), the non-perturbative part of the 4d 't Hooft line index, namely the monopole bubbling index, which naively should arise from { the large orbifold limit of the instanton part of the 5d index}, is yet to be fully understood.  This is intimately related to the fact that the 5d analogue of the monopole bubbling effect in 4d is still unclear.

The puzzle comes from the naive matching of parameters in the 5d instanton index with those in the 4d monopole bubbling index.
The 5d instanton index on an orbifold can depend only on the monodromy $\vec{r}$ at infinity.  From the perturbative part in the large orbifold limit, the vector $\vec r$ becomes the weight $\vec v$ in 4d, as can be seen from $I_p (\vec v)$ in \eref{eq:Ip} and \eref{tHooftindex}.  On the contrary, the $Z_{\rm mono}(\vec{B}, \vec{v})$ does depend on both weight $\vec{v}$ and the chosen representation $\vec{B}$.  This mismatch of the parameters lead us to speculate the following possibility to define the $5d$ analogue of the monopole bubbling.

For a 5d theory on $\mathbb{C}^2/\mathbb{Z}_n$, it is not enough with choosing one single monodromy, but we may need to sum over the whole set of other monodromies.\footnote{Note however that fixing one single monodromy also seems a consistent procedure. As a consistency check, upon choosing a single monodromy and reducing along the ``time" $S^1$ we find the 4d partition function on an ALE space, where no bubbling effect has been described in the literature.} Let us proceed along the same way as for the 4d 't Hooft line index \eref{tHooftindex}.  Take $\vec r$ to be a representation of $U(N)$ and take $\vec \rho$ to be a weight of $\vec r$.  We denote the set of weights of $\vec r$ by ${\cal W}_{\vec r}$.  We speculate that the 5d index reads
\begin{equation}
\label{5dindex}
\mathcal{I}_{\rm 5d} (\vec r, \vec \rho)=\sum_{\vec{\rho}\in{\cal W}_{\vec r}}\,\int [d \vec z]_{\vec{\rho}}\, \mathcal{I}_{\rm p}^{\mathbb{C}^2/\mathbb{Z}_n}(\vec{\rho})\,\mathcal{I}_{\rm np}^{\mathbb{C}^2/\mathbb{Z}_n}(\vec r, \vec{\rho})~.
\end{equation}
We interpret $\mathcal{I}_{\rm np}^{\mathbb{C}^2/\mathbb{Z}_n}$ as the instanton contribution in 5d, and so it should depend only on one monodromy at infinity; we take that to be $\vec{\rho}$. One natural guess is to write \eref{5dindex} as
\begin{equation}
\label{5dindexrewritten}
\mathcal{I}_{\rm 5d}(\vec{r}, \vec{\rho})=\sum_{\vec{\rho}\in {\cal W}_{\vec r} }\,\int [d\vec z]_{\vec{\rho}}\,\mathcal{I}_{\rm np}^{\mathbb{C}^2/\mathbb{Z}_n}(\vec{r})\, \mathcal{I}_{\rm p}^{\mathbb{C}^2/\mathbb{Z}_n}(\vec{\rho})\,\widehat{\mathcal{I}}_{\rm np}^{\mathbb{C}^2/\mathbb{Z}_n}(\vec{r}, \vec{\rho})\, , \qquad \widehat{\mathcal{I}}_{\rm np}^{\mathbb{C}^2/\mathbb{Z}_n}(\vec{r}, \vec{\rho})=\frac{\mathcal{I}_{\rm np}^{\mathbb{C}^2/\mathbb{Z}_n}(\vec{\rho})}{\mathcal{I}_{\rm np}^{\mathbb{C}^2/\mathbb{Z}_n}(\vec{r})}~.
\end{equation}
Interpreting the ``overall" $\mathcal{I}_{\rm np}^{\mathbb{C}^2/\mathbb{Z}_n}(\vec{r})$ as a ``Casimir energy''\footnote{Note it is not quite an overall factor, as it depends on gauge fugacities. Nevertheless, the gauge fugacity dependence also occurs in the quantity in \textit{e.g.} eq. (3.48) of \cite{Gang:2012yr}, where they have been set to one.}, dropping it we find

\begin{equation}
\label{5dindexguess}
\widehat{\mathcal{I}}_{\rm 5d} (\vec{r}, \vec{\rho})=\sum_{\vec{\rho}\in{\cal W}_{\vec r}}\,\int [d\vec z]_{\vec{\rho}}\, \mathcal{I}_{\rm p}^{\mathbb{C}^2/\mathbb{Z}_n}(\vec{\rho})\,\widehat{\mathcal{I}}_{\rm np}^{\mathbb{C}^2/\mathbb{Z}_n}(\vec{r}, \vec{\rho})~.
\end{equation}
It is then natural to conjecture that, in the large $n$ limit, this quantity becomes the 4d 't Hooft line index. While the perturbative part of this quantity, together with the measure, recovers the expected 4d perturbative result with the insertion of a 't Hooft line, we would conjecture that the quantity $\widehat{\mathcal{I}}_{\rm np}^{\mathbb{C}^2/\mathbb{Z}_{\infty}}(\vec{r}, \vec{\rho})$ becomes the monopole bubbling contribution. Note that this proposal automatically incorporates that, for the highest weight $\vec \rho = \vec r$ of the representation $\vec r$, $\widehat{\mathcal{I}}_{\rm np}^{\mathbb{C}^2/\mathbb{Z}_n}(\vec{r},\,\vec{r})=1$. Thus, we would identify the chosen representation $\vec{r}$ with $\vec{B}$ and the weights $\vec \rho$ with the weights $\vec v$ in the large orbifold limit. Note also that SUSY requires $\widehat{\mathcal{I}}_{\rm np}^{\mathbb{C}^2/\mathbb{Z}_{\infty}}(\vec{r}, \vec{\rho})$ to depend only on the product $t_1\,t_2$.

As a direct test of this proposal, we can consider the trivial monodromy case $\vec{r}=\vec{0}$ for an arbitrary 5d gauge theory. In this case, there is no sum and the 5d index is just the product of the perturbative and the instanton contribution. In the large $n$ limit, and dropping the non-perturbative contribution due to \eref{5dindexrewritten}, we just have the perturbative part, whose large orbifold limit we know to reproduce the Schur index. 

Note that another subtlety is the fact that, while the 5d instanton index will depend on the instanton fugacity $q$ \cite{Kim:2012gu} (see also (2.6) of \cite{Ito:2013kpa}). Upon taking the large orbifold limit we recover a 4d partition function \eref{tHooftindex} for which we do not expect such fugacity. Thus we expect that the large $n$ limit makes the explicit $q$-dependence to disappear.\footnote{One way this might happen is due to the fact that instanton numbers on $\mathbb{C}^2/\mathbb{Z}_n$ are multiples of $1/n$. Hence large $n$ is like effectively setting $q=1$.} It is instructive to consider the case of a 5d pure $U(1)$ gauge theory on an orbifold. Note that for the pure $U(1)$ gauge theory the orbifold cannot act on the gauge fugacities. The exact 5d index on $\mathbb{C}^2$ was computed in Eq. (25) of \cite{Rodriguez-Gomez:2013dpa}: 
\bea
\CI^{\BC^2}_{\text{inst}} = \PE \left[f^{\BC^2}_{\vect}  +  (q+q^{-1})f^{\BC^2}_{\text{adjoint}}  \right] = \PE \left[ \frac{-(t_1+t_2)}{(1-t_1)(1-t_2)}+\frac{\sqrt{t_1t_2} (q+q^{-1})}{(1-t_1)(1-t_2)}  \right]~,
\eea
where the first term corresponds to the perturbative part  and the second term corresponds to the instanton contributions from the north and the south poles.
Projecting it to orbifold-invariants, in the large orbifold limit, we find that it reads 
\bea 
\CI^{\BC^2/\BZ_n}_{\text{inst}} (t_1,t_2,q)
&=\PE \left[ \oint_{|w|=1} \frac{\ud w}{2 \pi i w}  \frac{-(wt_1+w^{-1}t_2)+\sqrt{t_1t_2} (q^{1/n}+q^{-1/n})}{(1-wt_1)(1-w^{-1}t_2)}   \right] \nn \\
&= {\rm PE} \left[-\frac{2t_1 t_2}{1-t_1t_2}+\frac{\,\sqrt{t_1t_2}}{1-t_1t_2} (q^{1/n}+q^{-1/n})  \right] \nn \\
&\sim   {\rm PE} \left[-\frac{2t_1 t_2}{1-t_1t_2}+\frac{2\sqrt{t_1t_2}}{1-t_1t_2} \right]~, \qquad n \rightarrow \infty~.
\eea
Note that the first and second terms are the Schur index of the 4d $U(1)$ vector field and that of an extra hypermultiplet, respectively. Our proposal amounts to dropping the second term, hence finding the Schur index of the pure $U(1)$ theory in 4d. 

It is not clear to us the underlying reason to our proposal. Note however that, as mentioned above, the large orbifold limit produces a singular space. It might well be that our procedure amounts to effectively remove the effect of such singularity. Indeed, in the pure $U(1)$ example above, it's tempting to identify the extra hyper with the produced singularity. 

Another salient result of our work is that, by studying in detail the monopole bubbling index we have found that it can be computed as the Hilbert series of a certain instanton moduli space. Such instanton lives on an orbifold whose degree is specified in a precise way by the charge $\vec B$ of the monopole which is being inserted, and whose modromy is given to the bubbling $\vec v$. We stress that this instanton on an orbifold construction is an auxiliary device which allows to easily compute monopole bubblings in the spirit of Kronheimer's construction  \cite{kronheimer1985monopoles}, and it should not be confused with the physical orbifold where the 5d theory lives on. 

One might also consider simultaneous reduction on the ``time" $S^1$ and the orbifolded circle. Naively this would lead to the partition function of the 3d version of the theory in the presence of a monopole operator. We leave for future work the study of this possibility. Note that 5d theories with $U(N)$ gauge group admit a 5d Chern-Simons term, whose effect will enter the instanton part of the 5d index. We leave for future work the study of the effect of such CS, in particular its effect in the reduced theories.

Lastly, we have mostly focused on the case of $U(N)$ gauge theories, but a similar analysis should be possible for other gauge groups. In particular, by carefully studying the possible actions of the orbifold on the gauge group should be equivalent, through the large orbifold limit, to the study of the allowed line defects for a given gauge group, hence giving detailed non-perturbative information about the global structure of the theory in question \cite{Aharony:2013hda, Razamat:2013opa}.

\section*{Acknowledgements}

We are grateful to Yuto Ito, Seok Kim, Kimyeong Lee and Takuya Okuda for very useful conversations. N.~M.~ would like to thank the Simons Center for Geometry and Physics for the hospitality and for providing the opportunity for fruitful discussions. D.~R-G.~ would like to thank the Centro de Ciencias Pedro Pascual de Benasque for hospitality while this work was in progress. The work of N.~M.~ is supported by a research grant of the Max Planck Society and is based upon work supported in part by the National Science Foundation under Grant No. PHY-1066293 and the hospitality of the Aspen Center
for Physics. D.~R-G.~ is partially supported by the research grants FPA2012-35043-C02-02, as well as by the Ramon y Cajal fellowship RyC-20011- 07593.
 
N.~M.~ also would like to use this opportunity to express his sincere gratitude to the Max-Planck-Institut f\"ur Physik 
(Werner-Heisenberg-Institut) for providing great facilities and nice environment for his research during the two years of his postdoctoral position.  He deeply thanks Thomas Grimm, the Group Leader, for his encouragement and very generous support.  N.~M.~ has greatly benefited from collaborating with Dieter L\"ust and from a number of discussions with his colleagues and friends.  It has been a privilege and a fantastic time for N.~M.~ to conduct his research in M\"unchen.
\begin{appendix}

\section{Monopole bubbling indices for $U(N)$ pure gauge theory}\label{largeB}

Using the correspondence found in section \ref{monopolebubbling} we can obtain monopole bubbling indices by computing Hilbert series for instanton moduli spaces following the techniques of \cite{DHMRS2013}. In this appendix we use such techniques to compute several exact results.

\subsection{$\vec B=(n,0^{N-1})$ and $\vec v=(1,n-1,0^{N-2})$}
These $\vec B$ and $\vec v$ correspond to the solution $\vec K =(1,2,3, \ldots, n-1)$ of \eref{eqKUN}.  From Section \ref{sec:monoVShs}, the monopole bubbling index is equal to the Hilbert series of $(n-1)/n$ pure $SU(N)$ instantons on $\BC^2/\BZ_n$, correpsonding to $\vec k=(1^{n-1},0)$ and $\vec r =(1,n-1,0^{N-2})$\footnote{The superscript indicates the number of repetitions.}.

As shown in \cite{DHMRS2013}, the moduli space of such instantons  is $\BC^2/\BZ_n$. Hence, the monopole bubbling index is
\bea
&Z_{\text{mono}} (\vec B=(n,0^{N-1}), \vec v=(1,n-1,0^{N-2})) (x, \vec z)  \nn \\
&= g_{\BC^2/\BZ_n} (x, z) \nn \\
&= \PE[x^2+ x^{n} (z^n +z^{-n})  - x^{2n} ] \nn \\
&= \frac{1-x^{2n}}{(1- x^n z^n)(1- x^n z^{-n})(1-x^2)}~,
\eea
where $g_{\BC^2/\BZ_n} (x, z)$ is the Hilbert series of $\BC^2/\BZ_n$.  In fact, we have seen special cases of this for $N=2$, $n=2$ in \eref{Zmono2011pureU2}, for $N=2$, $n=3$ in \eref{HSC2Z3} and for $N=3$, $n=2$ in \tref{tab:matchU3SYM}.

In the large $n$ limit, the index reduces to the Hilbert series of $\BC$:
\bea
Z_{\text{mono}}( \vec B = (n,0^{N-1}), \vec v =(1,n-1,0^{N-2})) (x) \sim \frac{1}{1-x}~, \qquad n \rightarrow \infty~.
\eea

\subsection{$\vec B=(1,-1,0^{N-2})$ and $\vec v=(0^{N})$}
These $\vec B$ and $\vec v$ correspond to the solution $\vec K =(0)$ of \eref{eqKUN}.  From Section \ref{sec:monoVShs}, the required monopole bubbling index is equal to the Hilbert series of $SU(N)$ instantons on $\BC^2/\BZ_n$, with $\vec k=(0^{n-1},1)$ and $\vec r =(0^{N})$ or $\vec N=(0^{n-1},N)$.

As shown in \cite{DHMRS2013, Benvenuti:2010pq}, the moduli space of such instantons is equal to the moduli space of one $SU(N)$ instanton on $\BC^2$, whose Hilbert series is given by \cite{Benvenuti:2010pq}. Explicitly,
\bea
Z_{\text{mono}}(\vec B=(1,-1, 0^{N-2}), \vec v =(0^{N}))(x, \vec z) =  \sum_{m=0}^\infty [m,0,\ldots,0,m]_{\vec z} x^{2m}~,
\eea
where $[1,0,\ldots,0,1]_{\vec z}$ denotes the character of the adjoint representation of $SU(N)$.

\subsection{$U(2)$ theory with $\vec B=(n,0)$ and $\vec v=(p,n-p)$}
These $\vec B$ and $\vec v$ correspond to the following solution of \eref{eqKUN}.
\bea 
\vec K &=(1^1,2^2,3^3, \ldots,p^p, (p+1)^{p}, (p+2)^{p}, \ldots, (n-p)^p,  \nn \\
& \qquad (n-p+1)^{p-1}, \ldots, (n-3)^3,(n-2)^2,(n-1)^1) ~.
\eea
Hence, the required monopole bubbling index is equal to the Hilbert series of pure instantons with
\bea
\vec r =(p,n-p), \quad \vec k =(1,2,3, \ldots,p-1, p^{n-2p+1},p-1, \ldots,1,0, \ldots,0)~; 
\eea
this corresponds to the instanton number $k=p(n-p)/n$.  Explicit expressions for Hilbert series for certain $(n,p)$ can be found, \eg~ \eref{B40v31} for $(4,1)$ and \eref{B40v22} for $(4,2)$.

\paragraph{The large orbifold limit.} Let us consider the limit $n \rightarrow \infty$.  For $\vec r=(p, -p)$, with $p \geq 0$, we find that the Hilbert series for such instantons, or equivalently the required monopole bubbling index is
\bea
& Z_{\text{mono}} (\vec B=(n,0), \vec v=(p,n-p)) (t_1,t_2) \nn \\
&\sim \prod_{m=1}^p \frac{1}{1-(t_1 t_2)^m} \label{inforbksmallestinfprod} = \PE \left[ \sum_{m=1}^p (t_1 t_2)^m \right] = \PE \left[  \frac{t_1 t_2}{1-t_1 t_2} \left \{ 1-(t_1t_2)^p \right \} \right]~, \qquad n \rightarrow \infty
\eea
Note that there is no dependence on $\vec z$ in the large orbifold limit.

\end{appendix}

\bibliographystyle{ytphys}
\bibliography{ref}

\end{document}